\documentclass{aa}
\usepackage{graphicx}
\usepackage{txfonts}
\usepackage{amsmath}
\usepackage{amsfonts}

\def\kms{km~s$^{-1}$}
\def\teff{$\rm T_{\rm eff}$}

\def\gr{$\log {\rm g}$}
\def\vt{${\rm v_{\rm t}}$}
\def\gk{$(\rm G-K_{s})_{0}$}
\def\ks{${\rm K_{s}}$}

\begin{document}

\title{The chemical DNA of the Magellanic Clouds}
\subtitle{I. The chemical composition of 206 Small Magellanic Cloud red giant stars
\thanks{Based on observations collected at the ESO-VLT under the programs 072.D-0507, 083.D-0208 and 086.D-0665.}}

\author{
A. Mucciarelli\inst{1,2} \and
A. Minelli\inst{1,2} \and 
M. Bellazzini\inst{2} \and
C. Lardo\inst{1} \and
D. Romano\inst{2} \and
L. Origlia\inst{2} \and
F. R. Ferraro\inst{1,2}
}

\institute{
Dipartimento  di  Fisica  e  Astronomia  “Augusto  Righi”,  Alma  Mater  Studiorum, Universit\`a  di Bologna, Via Gobetti 93/2, I-40129 Bologna, Italy
\and
INAF - Osservatorio di Astrofisica e Scienza dello Spazio di Bologna, Via Gobetti 93/3, I-40129 Bologna, Italy
}

\authorrunning{Mucciarelli et al.}
\titlerunning{SMC field stars}

\abstract{
We present the chemical composition of 206 red giant branch stars members of the Small Magellanic Cloud (SMC) using optical, 
high-resolution spectra collected with the multi-object spectrograph FLAMES-GIRAFFE at the ESO Very Large Telescope. 
This sample includes stars in three fields located in different positions within the parent galaxy. 
We analysed the main groups of elements, namely light- (Na), $\alpha$- (O, Mg, Si, Ca, Ti), iron-peak (Sc, V, Fe, Ni, Cu) 
and s-process elements (Zr, Ba, La). The metallicity distribution of the sample
displays a main peak around [Fe/H]$\sim$--1 dex and a weak metal-poor tail.
However, the three fields display [Fe/H] distributions different with each other, in particular a difference of 0.2 dex 
is found between the mean metallicities of the two most internal fields.
The fraction of metal-poor stars increases significantly (from $\sim$1 to $\sim$20\%) from the innermost fields to the most 
external one, likely reflecting an age gradient in the SMC. Also, we found a hint of possible 
chemically/kinematic distinct substructures. 
The SMC stars have abundance ratios clearly distinct with respect to the Milky Way stars, in particular for the elements produced 
by massive stars 
(like Na, $\alpha$ and most iron-peak elements) that have abundance ratios systematically lower than those measured in 
our Galaxy. This points out that 
the massive stars contributed less to the chemical enrichment of the SMC 
with respect to the Milky Way, according to the low star formation rate 
expected for this galaxy. Finally, we identified small systematic differences 
in the abundances of some elements (Na, Ti, V and Zr) in the two innermost fields, suggesting that the chemical enrichment history 
in the SMC has been not uniform.}
 
              
\maketitle

\section{Introduction}

The Local Universe provides an unique window into the process of hierarchical mass assembly on all scales, 
allowing us to investigate a plethora of systems, all of them satellites of the major assemblies, like the Milky Way (MW) 
and M33: for instance, galaxies in relative isolation (like most of the nearby dwarf galaxies), 
in close interaction with other systems (the Large and Small Magellanic Cloud, LMC and SMC, respectively) or 
consumed by large galaxies (like the Sagittarius dwarf remnant and 
the satellites engulfed by the MW). 
Among them, the Magellanic Clouds, thanks to their proximity and the possibility to resolve individual stars, 
provide an unique close-up of a pair of interacting dwarf galaxies.

They are gas-rich irregular galaxies, gravitationally bound each other and likely
at the first peri-Galactic passage with the MW \citep{besla07,besla10,kalli13,besla15}. 
The galaxy discussed in this paper, the SMC, is the second most massive MW satellite after the LMC, 
with a total mass of $\sim2\cdot10^{9}M_{\odot}$ \citep{stan04}, about one order of magnitude lower than that of the LMC, 
and a stellar mass of $\sim5-6\cdot10^{8}M_{\odot}$ \citep{vdm09,rubele18}, comparable with that of 
the main merger of the Milky Way, the former galaxy Gaia-Enceladus \citep{helmi18}. 
There are several signatures of their mutual 
interaction and of the interaction between the Clouds and the MW, like the Magellanic Bridge connecting SMC and LMC and 
the Magellanic Stream embracing these two galaxies.
The history of the stellar populations of the SMC is intimately linked to the interplay of these three galaxies \citep{massana22}: 
the multiple episodes of star formation (SF) occurring in their history are likely the result of the periodic 
close encounters between them.
The colour-magnitude diagrams (CMD) of different SMC fields \citep[see e.g.][]{harris04,noel07,cignoni12,cignoni13} 
reveal a mixture of stellar populations, with a prominent red giant branch (RGB), 
signature of stellar populations older than 1-2 Gyr, and the presence of an extended 
blue main sequence, hinting at the presence of younger stars. 
Our current picture of the SMC SF history \citep{cignoni12,cignoni13,rubele18,massana22} is that this galaxy 
formed in isolation, with a SF activity starting $\sim$13 Gyr ago and 
a prolonged period of low-level SF activity until $\sim$3-4 Gyr ago. At this epoch, the SMC has been 
likely tidally captured by the LMC, becoming gravitationally bound to it. This capture should have
triggered new, vigorous and synchronised SF bursts in both the galaxies \citep[see e.g.][]{bekki05,massana22}, 
likely forming most of the stars that we observe today. 
According to \citet{rubele18}, the SMC formed (5.31$\pm$0.05)$\times 10^{8} M_{\odot}$
of stars over an Hubble time, 2/3 of which are now found in stellar remnants or living stars.\\
At variance with the LMC stars whose chemical composition has been widely studied using high resolution spectroscopy \citep{hill00,pompeia08,muc10,lapenna12,vds13,nidever20,muc21}, the chemical composition of the SMC stars 
has received less attention, despite the proximity of this galaxy \citep[$\sim$62 kpc,][]{graczyk14}.\\
For decades, the only high-resolution spectroscopic studies of SMC stars were mainly focused 
on bright supergiant stars and cepheids, hence sampling stellar populations younger than $\sim$200 Myr
\citep[see e.g.][]{spite89a,spite89b,hill97,romaniello08}. 
Most of the information about the metallicity distribution of the SMC RGB stars 
came from low-resolution spectroscopy in I-band, using the calibrated strength 
of the Ca~II triplet as a proxy of [Fe/H] \citep{carrera08,dobbie14a,dobbie14b,parisi16,deleo20}.
The metallicity distribution of the SMC stars as derived from these studies displays a 
main peak around [Fe/H]$\sim$--1 dex and a weak metal-poor tail. A clear decrease 
of the mean metallicity has been observed at distance larger than $\sim3^{\circ}$ from the galaxy centre \citep{carrera08}.
Also, evidence of a shallow metallicity gradient within the SMC's inner $\sim$3$^{\circ}$ have been found, 
between --0.07 dex/deg \citep{dobbie14a} and --0.03 dex/deg \citep{chou2020}.

Only recently, chemical analyses of high-resolution spectra of SMC RGB stars have been 
presented \citep{nidever20,reggiani21,hasselquist21}, allowing us to investigate in details 
the chemical composition of these stellar populations. \citet{nidever20} discussed [Mg/Fe], [Si/Fe] and [Ca/Fe] abundance 
ratios for about 1000 RGB SMC stars, finding a quite flat behaviour of these abundance ratios in the range of [Fe/H] between 
--1.2 and --0.2 dex, and a {\sl knee} (the metallicity corresponding to the decrease of the [$\alpha$/Fe] abundance ratios) 
located at [Fe/H] lower than --2.2 dex. 
The same sample of SMC stars is discussed by \citet{hasselquist21} that 
includes also the abundances of Al, O, Ni and Ce, and compare them with those of other Milky Way satellites. 

\citet{reggiani21} discussed the chemical composition of four metal-poor ([Fe/H]$<$--2.0 dex) SMC stars, finding 
that these stars have abundances comparable to those of the MW halo stars for all the main groups of elements. 
On the other hand, these stars are more enriched in [Eu/Fe] (a pure r-process element) with respect to the MW stars.

This paper is the first of a series dedicated to the investigation of the chemical properties of the LMC/SMC (field and 
stellar clusters) stars. In this work, we present the chemical analysis of 206 RGB stars members of the SMC observed 
with the high-resolution spectrograph FLAMES mounted at the ESO Very Large Telescope.  


\section{Observations and data reduction}

\subsection{SMC sample}
\begin{figure*}[htbp]
\includegraphics[clip=true,width=1\textwidth]{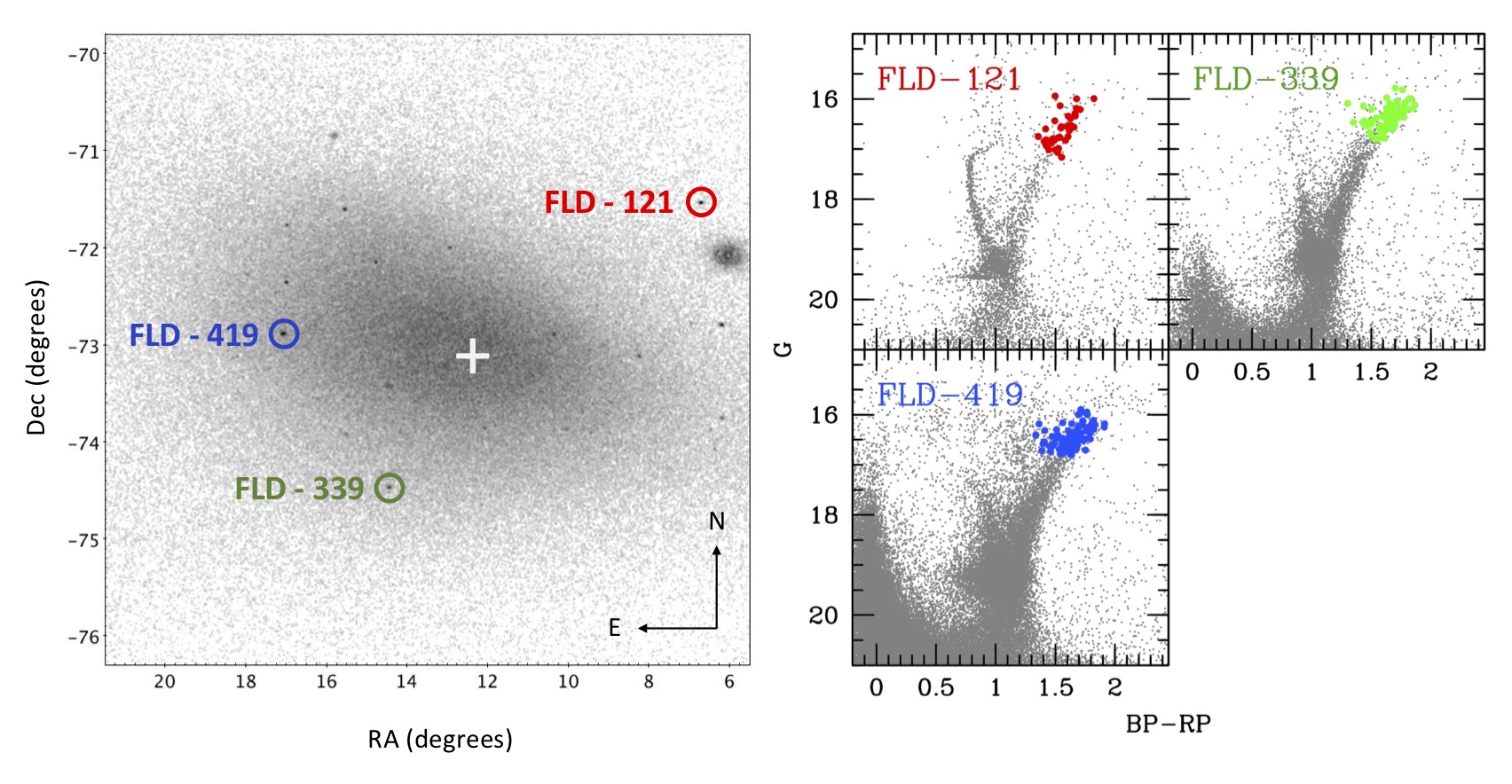}
\caption{Left panel: spatial distribution of the three fields 
observed with FLAMES (marked as red, green and blue circles for FLD-121, FLD-339 and FLD-419, respectively) 
superimposed to the map of the SMC RGB stars with G between 16 and 19 from Gaia EDR3 \citep{brown21}, revealing 
the SMC old spheroid.
The white plus symbol marks the position of the SMC centre derived 
by \citet{ripepi17}. 
Right panel: Gaia EDR3 CMDs of the three SMC fields \citep[grey points, ][]{brown21} with superimposed the spectroscopic 
GIRAFFE targets (same colours of left panel).
In the CMD of FLD-121 is clearly visible the main sequence of the MW GC 47 Tucanae.
}
\label{map}
\end{figure*}

A total of 320 stars in the direction of the SMC has been observed (ID program 086.D-0665, PI: Mucciarelli) 
with the multi-object spectrograph FLAMES \citep{pasquini} in the GIRAFFE-MEDUSA mode that allows us the simultaneous 
allocation of 132 high-resolution (R$\sim$20,000) fibres over a patrol field of about 25 arcmin diameter.
Three different fields have been observed, centred around three globular clusters (GCs),
namely NGC~121, NGC~339 and NGC~419 (hereafter these fields will be indicated as 
FLD-121, FLD-339 and FLD-419, respectively). 
Left panel of Fig.~\ref{map} shows the spatial location of the three FLAMES fields superimposed to the 
map of the SMC stars obtained with the early third data release (EDR3) of the {\sl Gaia}/ESA mission
\citep{Gaia16,brown21}. `
The fields are located in different positions of the SMC, with 
FLD-121, FLD-339 and FLD-419 at $\sim2.4^{\circ}$ northern-western, $\sim1.4^{\circ}$ southern-eastern 
and $\sim1.5^{\circ}$ eastern from the SMC centre \citep{ripepi17}, respectively.
The field FLD-121 partially overlaps with the APOGEE field 47Tuc 
(two only stars in common), the field FLD-419 is adjacent to the APOGEE field SMC2 (one only star 
in common), while the field FLD-339 samples a region not observed by APOGEE \citep[see Fig. 1 by ][]{nidever20}.

The adopted GIRAFFE-MEDUSA setups are HR11, with a spectral resolution of 24200 and ranging from 5597 
to 5840 \AA\, and HR13, with a spectral resolution of 22500 and a spectral coverage between 6120 and 6405 \AA\ . 
These two setups allow us to measure lines of the main groups of elements, like odd-Z (Na),  $\alpha$ (O, Mg, Si, Ca and Ti), 
iron-peak (Sc, V, Fe, Ni, Cu) and s-process elements (Zr, Ba, La).
The UVES fibers have been allocated to targets belonging to the three GCs and discussed in separated papers 
\citep[][A. Minelli et al., in prep.]{dalex16}.
Table~\ref{tab1} lists the exposure times and the number of individual exposures for each setup and field.


\begin{table*}
\caption{Coordinates of the FLAMES pointing,  number of exposures and exposure times 
for the two FLAMES setups, adopted colour excess \citep{schlafly11} and the number 
of SMC stars analysed in this work.}            
\label{tab1}     
\centering                        
\begin{tabular}{lcccccc}        
\hline\hline                 
Field &   RA & Dec  & HR11& HR13  & E(B-V) & $N_{SMC}$   \\ \\    
\hline
  &   (J2000)  &  (J2000)  &  &  & (mag)  &     \\
\hline   
FLD-121 & 00:26:49.0   &  --71:32:09.9  &  7x2700sec     &  5x2700sec  &  0.028   & 37  \\    
        &              &                &  1x2200sec     &             &          &     \\	  
FLD-339 & 00:57:48.9   &  --74:28:00.1  &  9x2700sec     &  5x2700sec  &  0.042   & 78  \\  
FLD-419 & 01:08:17.7   &  --72:53:02.7  &  6x2700sec     &  4x2700sec  &  0.089   & 91 \\   
\hline
\end{tabular}
\end{table*}

The spectroscopic targets for each field have been originally selected from  near-infrared ($K_s$,J-\ks) 
CMDs, using the SofI@NTT catalogues for the region within 2.5 arcmin from the cluster centres
\citep[][for NGC~339 and NGC~419, and unpublished proprietary photometry for NGC~121]{m09}, 
and the 2MASS database \citep{skrutskie} for the external regions.
The targets have been selected according to the following criteria:
(1)~stars fainter than the RGB Tip \citep[\ks=12.62,][]{cioni00}; 
(2)~stars brighter than \ks=~14 for FLD-339 and FLD-419, and 
brighter than \ks=~14.4 for FLD-121,
in order to guarantee a  signal-to-noise ratio (SNR) per pixel larger than $\sim$30 in both setups and in all the 
observed fields. 
Due to the paucity of RGB stars in the SMC outskirts, 
a fainter (by $\sim$0.4  mag) threshold has been adopted for FLD-121 in order to enlarge the number of observed SMC stars;
(3)~stars without close stars brighter than  $<K_{s}^{star}+$1.0 within 2'' ;
(4)~for the targets from the 2MASS catalogue (the majority of the observed targets) 
only stars with J and \ks\ magnitudes flagged as {\sl A} (photometric uncertainties 
smaller than 10\%) have been selected.

All the targets have been recovered in the {\sl Gaia} EDR3 catalog. 
Right panel of Fig.~\ref{map} shows the position in the (G, BP-RP) CMDs of the observed targets considered as SMC stars 
according to their radial velocity (RV), see Section~\ref{rvd}. 
Table~\ref{stars} lists for all the SMC targets coordinates and the Gaia EDR3 identification number.

\begin{table*}
\caption{Information about the SMC spectroscopic targets: ID for our internal catalogues, ID  and coordinates 
from Gaia EDR3 \citep{brown21}, measured RV, derived atmospheric parameters and [Fe/H] abundance ratio. 
The entire table is available in electronic form.}            
\label{stars}     
\centering                 
\begin{tabular}{c ccc  c ccc c}        
\hline
 ID              &     ID Gaia EDR3        &     RA        &     Dec         & RV  &  \teff\ & \gr\  & \vt\  &  [Fe/H] \\ 
\hline  
                 &                         &   (degree)    &  (degree)       & (\kms )	& (K)  & (cgs) & (\kms ) &	(dex)  \\	 
\hline
FLD-121\_23      &  4689857932203222528    &   6.6427941   &  -71.5293047   &   144.3$\pm$ 0.2 &   4115  &   0.79  &  1.8  & --1.58$\pm$ 0.10  \\ 
FLD-121\_24      &  4689857863486443520    &   6.6292441   &  -71.5407885   &   146.9$\pm$ 0.1 &   4140  &   0.80  &  1.8  & --1.40$\pm$ 0.11  \\ 
FLD-121\_50      &  4689845798923576704    &   6.7124313   &  -71.5774614   &   123.5$\pm$ 0.1 &   4319  &   1.06  &  1.7  & --1.01$\pm$ 0.13  \\ 
FLD-121\_51      &  4689857691687749888    &   6.6839449   &  -71.5429782   &   150.8$\pm$ 0.3 &   4234  &   1.05  &  1.7  & --1.56$\pm$ 0.14  \\ 
FLD-121\_100004  &  4689848036601043584    &   7.0469194   &  -71.4811913   &   123.6$\pm$ 0.1 &   4142  &   0.80  &  1.8  & --0.93$\pm$ 0.10  \\ 
FLD-121\_100086  &  4689845597059733248    &   6.7120893   &  -71.5895200   &   106.2$\pm$ 0.1 &   4065  &   0.84  &  1.8  & --0.89$\pm$ 0.11  \\ 
FLD-121\_100175  &  4689859787631565056    &   7.0625549   &  -71.4769602   &   121.1$\pm$ 0.2 &   4375  &   0.98  &  1.7  & --1.32$\pm$ 0.13  \\ 
FLD-121\_100185  &  4689858172724094720    &   6.4880274   &  -71.5481267   &   140.2$\pm$ 0.1 &   4084  &   0.88  &  1.8  & --1.17$\pm$ 0.10  \\ 
FLD-121\_100211  &  4689852189834915072    &   6.2466252   &  -71.5899642   &   133.0$\pm$ 0.2 &   4345  &   1.09  &  1.7  & --1.14$\pm$ 0.14  \\ 
FLD-121\_100237  &  4689844978584591104    &   6.4954253   &  -71.6527985   &   150.8$\pm$ 0.1 &   4293  &   1.12  &  1.7  & --0.82$\pm$ 0.13  \\ 
FLD-121\_100263  &  4689843363676630528    &   7.1153070   &  -71.6019383   &   137.3$\pm$ 0.1 &   4132  &   0.84  &  1.8  & --1.01$\pm$ 0.10  \\ 
FLD-121\_100272  &  4689846730930989440    &   6.9812926   &  -71.5614137   &   170.8$\pm$ 0.1 &   4029  &   0.63  &  1.8  & --0.98$\pm$ 0.10  \\ 
FLD-121\_100330  &  4689859031717528576    &   6.5997764   &  -71.4819327   &   139.3$\pm$ 0.3 &   4424  &   1.09  &  1.7  & --1.75$\pm$ 0.17  \\ 
FLD-121\_100335  &  4689862914367958144    &   6.4217566   &  -71.4335678   &   131.3$\pm$ 0.1 &   4194  &   0.93  &  1.7  & --0.99$\pm$ 0.13  \\ 
FLD-121\_100365  &  4689851266416720768    &   6.2102084   &  -71.6388781   &   145.3$\pm$ 0.1 &   4093  &   0.63  &  1.8  & --1.09$\pm$ 0.10  \\ 
FLD-121\_100382  &  4689842569108117888    &   7.1356396   &  -71.6152387   &   121.9$\pm$ 0.1 &   4088  &   0.70  &  1.8  & --1.24$\pm$ 0.11  \\ 
FLD-121\_100440  &  4689842161085790080    &   7.0699869   &  -71.6558040   &   129.8$\pm$ 0.3 &   4488  &   1.14  &  1.7  & --1.25$\pm$ 0.17  \\ 
\hline                                   
\end{tabular}
\end{table*}

\begin{figure}[h]
\includegraphics[width=\hsize,clip=true]{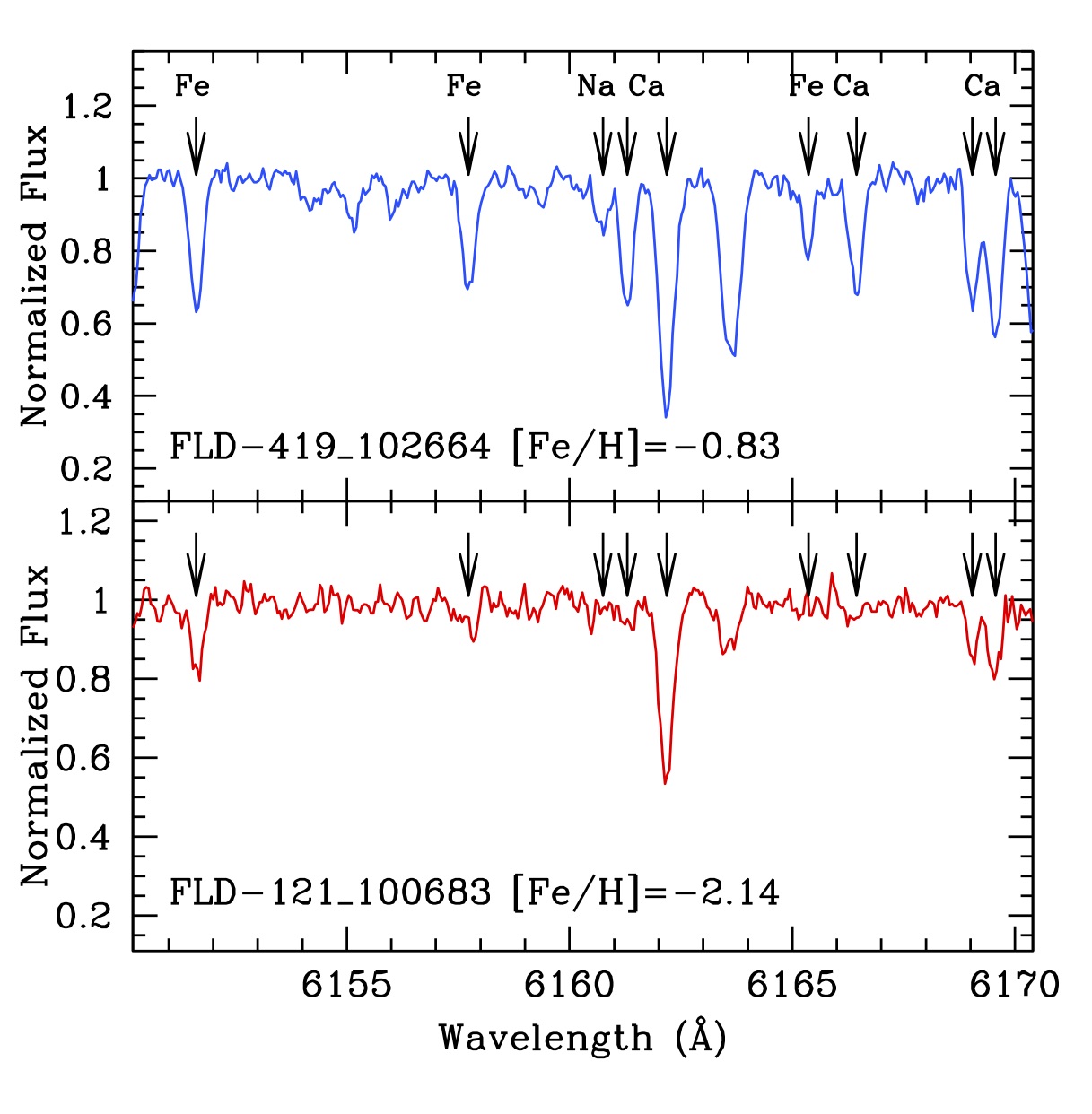}
\caption{Comparison between the HR13 spectra of the stars FLD-419\_102664 (upper panel) 
and FLD-121\_100683 (lower panel). The stars have very similar atmospheric parameters but 
different iron content. Arrows mark the position of some metallic lines of interest.}
\label{spec}
\end{figure}

The spectra have been reduced with the dedicated ESO GIRAFFE pipeline\footnote{https://www.eso.org/sci/software/pipelines/}, 
including bias-subtraction, flat-fielding, wavelength calibration with a standard Th-Ar lamp and spectral extraction. 
The contribution of the sky has been subtracted from each spectrum by using a median sky spectrum,
as obtained by combining $\sim$15-20 spectra from fibres allocated to sky positions within each exposure. 
The final SNR per pixel of the spectra is of $\sim$30-50 for HR11 spectra and $\sim$40-60 for HR13 spectra.
Fig.~\ref{spec} shows, as an example of the spectral quality, the spectra of two SMC giant stars with very similar 
atmospheric parameters and a large ($\sim$1.5 dex) difference in [Fe/H].


\subsection{MW control sample}

As discussed in \citet{minelli21}, the comparison between chemical abundances obtained from different 
works can be hampered by various systematics characterising 
the chemical analyses, for instance the method used to infer the stellar parameters, 
the adopted atomic data for the analysed transitions, model atmospheres and solar reference abundances. 
For this reason, when chemical analyses of extra-galactic stars are performed (with the aim 
to compare their abundances with those of MW stars), 
it is crucial to consider also a control sample of MW stars analysed in an homogeneous way, in order to 
erase the main systematics quoted above and highlight and quantify possible differences and similarities between the abundance 
ratios of stars from different galaxies.

We defined a control sample of MW stars analysed with the same assumptions used for the SMC stars.
We analysed five MW GCs covering the same metallicity range of the SMC stars ([Fe/H] 
between $\sim$--2.2 and $\sim$--0.5 dex) and for which FLAMES spectra obtained with the GIRAFFE HR11 and 
HR13 setups are available in the ESO archive (ID programs: 072.D-0507 and 083.D-0208, PI: Carretta). 
The selected GCs are NGC~104, NGC~1851, NGC~1904, NGC~4833 and NGC~5904. 
The use of the same GIRAFFE setups allows us to 
derive chemical abundances in these MW GCs from the same transitions used for the SMC stars.
We restrict the analysis only to the stars with effective temperatures and surface gravities comparable 
with those of the SMC stars studied here 
six stars for NGC~104, 2 for NGC~1851, 6 for NGC~5904, 3 for NGC~1904 and 4 for NGC~4833.
Only for O and Na that exhibit large star-to-star variations 
in each of these GCs, we analysed stars belonging to the so-called first population and selected 
according to \citet{carretta09}. 
The O and Na abundances of these first population stars can be considered as a good proxy 
of the chemical composition of the MW field at those metallicities.

\section{Spectral analysis}

\subsection{Line selection}

We selected for each star an appropriate set of unblended metallic lines, selected by visual inspection 
of suitable synthetic spectra. The latter have been calculated with the code {\tt SYNTHE} \citep{sbordone04,kurucz05},  
using the typical atmospheric parameters 
of the observed stars (see Section~\ref{atm}), adopting ATLAS9 model atmospheres 
\citep{castelli04}\footnote{http://www.oact.inaf.it/castelli/castelli/sources/atlas9codes.html} 
and including all the atomic and molecular transitions in the Kurucz/Castelli
linelist\footnote{http://www.oact.inaf.it/castelli/castelli/linelists.html}.
The synthetic spectra have been convoluted with Gaussian profiles in order to 
reproduce the observed line broadening, mainly dominated by the instrumental resolution.
We privileged transitions with laboratory oscillator strengths. 
Only for the Sc~II line at 6245.6 \AA\  , for the Si~I lines at 
6155.1 and 6237.3 \AA\ , and for the Cu~I line at 5782 \AA\ we adopted solar oscillator strengths. 

Because the level of blending of a given transition depends on the metallicity, in this case 
not known a priori, we adopted an iterative process to define the linelist of each target.

A preliminary linelist has been defined by adopting a metallicity [M/H]=--1.0 dex 
for all the used synthetic spectra, according to the mean metallicity of the 
SMC derived from previous studies \citep{carrera08,dobbie14a,dobbie14b,parisi16,nidever20}.
After a first chemical analysis, a new set of unblended lines has been defined for each star using a synthetic 
spectrum calculated with the appropriate chemical composition. 
This procedure has been specifically necessary for the most metal-poor stars of our sample, with [Fe/H] significantly 
lower than the mean value of [Fe/H]=--1.0 dex, and for a few stars with enhancement of s-process elements.
The average number of selected metallic lines is of about 80-90 for most of the stars (with [Fe/H]$\sim$--1.0 dex), 
decreasing down to 40-50 for the most metal-poor ones. 
Most of the lines used in the metal-poor stars are still available for metal-rich stars, while 
some features are excluded because saturated or blended with other lines at higher metallicities. 
However, we checked that the use of different samples of lines depending on the stellar metallicity 
does not introduce biases in the abundances at different metallicities.

All the used lines are listed in Table~\ref{atomic} together with the corresponding 
log~gf and excitation potential $\chi$.

\begin{table*}
\caption{List of the used transitions together with oscillator strengths, 
the excitation potential and the reference of the atomic data. 
Wavelengths without some decimal digits indicate transitions 
affected by hyperfine/isotopic splitting.
The entire table is available in electronic form.}            
\label{atomic}     
\centering                 
\begin{tabular}{c   c c c  c c}        
\hline\hline                 
Wavelength &   Ion &  loggf & $\chi$  & Reference  \\
\hline
\centering                 
 (\AA\ )   &     &    &  (eV)  &  \\
\hline   
5590.720  &  Co I     &  -1.870  &    2.042   &    \citet{fmw}       &    \\ 
5598.480  &  Fe I     &  -0.087  &    2.521   &    \citet{fw06}      &    \\ 
5601.277  &  Ca I     &  -0.523  &    2.526   &    \citet{sr81}      &    \\ 
5611.356  &  Fe I     &  -2.990  &    3.635   &    \citet{fmw}       &    \\ 
5615.644  &  Fe I     &   0.050  &    3.332   &    \citet{fw06}      &    \\ 
5618.632  &  Fe I     &  -1.276  &    4.209   &    \citet{fw06}      &    \\ 
5624.542  &  Fe I     &  -0.755  &    3.417   &    \citet{fw06}      &    \\ 
5633.946  &  Fe I     &  -0.320  &    4.991   &    \citet{fw06}      &    \\ 
5638.262  &  Fe I     &  -0.840  &    4.220   &    \citet{fw06}      &    \\ 
5647.234  &  Co I     &  -1.560  &    2.280   &    \citet{fmw}       &    \\ 
5648.565  &  Ti I     &  -0.260  &    2.495   &    \citet{mfw}       &    \\ 
5650.689  &  Fe I     &  -0.960  &    5.085   &    \citet{fw06}      &    \\ 
5651.469  &  Fe I     &  -2.000  &    4.473   &    \citet{fmw}       &    \\ 
5652.318  &  Fe I     &  -1.920  &    4.260   &    \citet{fw06}      &    \\ 
5653.867  &  Fe I     &  -1.610  &    4.386   &    \citet{fw06}      &    \\ 
5661.345  &  Fe I     &  -1.756  &    4.284   &    \citet{fw06}      &    \\ 
5662.516  &  Fe I     &  -0.573  &    4.178   &    \citet{fw06}      &    \\ 
5670.8**  &  V I      &  -0.420  &    1.081   &    \citet{mfw}       &    \\ 
5679.023  &  Fe I     &  -0.900  &    4.652   &    \citet{fw06}      &    \\ 
5682.633  &  Na I     &  -0.706  &    2.102   &    NIST              &    \\ 
\hline                                   
\end{tabular}
\end{table*}

\subsection{Atmospheric parameters}
\label{atm}

The derived atmospheric parameters are listed in Table~\ref{stars}.
Effective temperatures (\teff ) and surface gravities (\gr ) have been estimated from the photometry. 
In particular, \teff\ have been obtained from the broad-band colour \gk\ adopting 
the \gk\--\teff\ transformation provided by \citet{mbm21}. 
We adopted G magnitudes from {\sl Gaia} EDR3 and \ks\ from 2MASS.
G magnitudes have been corrected for extinction following the prescriptions by \citet{bab18}, 
while \ks\ magnitudes adopting the extinction coefficient by \citet{mccall04}.
The colour excess values E(B-V) are from the infrared dust maps by \citet{schlafly11} and listed in Table~\ref{tab1}.

Uncertainties in \teff\ have been estimated by propagating for any individual star 
the errors in the adopted colour and in the colour excess. 
The typical error in the \gk colours is of about 0.03-0.05 mag, dominated by the uncertainty of the \ks\ magnitude, 
and translating in 20-40 K of uncertainty in \teff\ . For the colour excess, we adopted a conservative error of 0.01 mag 
for all the three fields, despite the lower errors quoted by \citet{schlafly11}, leading to a negligible (a few K) 
uncertainty in \teff\ .
These uncertainties have been added in quadrature to the typical error associated to 
the \gk\--\teff\ transformation (46 K), estimated as 1$\sigma$ dispersion of the fit residuals \citep{mbm21}, 
and that dominates the total \teff\ error (typically $\sim$50-60 K).

The \gr\ values have been calculated through the Stefan-Boltzmann relation 
adopting the photometric \teff, a true distance modulus $(m-M)_0$=18.965$\pm$0.025 
\citep{graczyk14}, the bolometric corrections by \citet{andrae18} and a stellar mass 
of 1.0$M_{\odot}$. 
Concerning the distance modulus, it is worth to recall that
the SMC has a substantial 
line-of-sight depth, not easy to properly take into account for each individual target. 
According to the depth maps provided by \citet{sub09}, the three fields studied in this study should cover 
a depth range between 2 and 6 kpc. Considering a conservative variation 
of the distance by 3 kpc increases the quoted uncertainties in log~g only by 0.02 dex, 
translating in variations of less  
of less than 0.02  in the abundances of single ionised (but without impact on the abundances 
of neutral lines).
Uncertainties in log~g are of the order of 0.1, including 
the uncertainties in \teff\, distance modulus and stellar mass. 
The final error budget in log~g is dominated by the uncertainty in the stellar mass, assumed to be $\pm$0.2 $M_{\odot}$, 
reflecting the possible spread in ages of our targets (older than $\sim $1-2 Gyr).
Microturbulent velocities (\vt ) are usually derived spectroscopically by erasing any trend 
between iron abundance and the reduced equivalent widths (defined as the logarithm of the EW normalised to the wavelength). 
Because of the relatively small number ($\sim$30-40 or less) of available Fe~I lines in the adopted spectral ranges, 
\vt\ obtained spectroscopically risk to be uncertain or unreliable. 
In order to avoid significant fluctuations in  \vt\ 
(with an impact on the derived abundances), we adopted the \gr\ - \vt\ relations provided by \citet{mpb20} 
and based on the spectroscopic \vt\ obtained from high-resolution, high-SNR spectra of giant stars 
in 16 Galactic GCs. The uncertainty in \vt\ has been estimated by adding in quadrature the 
error arising from the uncertainty in \gr\ and that of the adopted \gr\ - \vt\ 
relation and is of the order of 0.2 \kms\ .

\subsection{Radial velocities}
\label{rvd}

RVs have been measured by using the code {\tt DAOSPEC} \citep{stetson08} that performs a line fitting 
assuming a Gaussian profile. The code is automatically launched by using the software {\tt 4DAO} \citep{4dao} 
that allows us a visual inspection of all the fitted lines in order to directly evaluate the quality of the fitting 
procedure. RVs have been measured by the position of about 100 metallic lines for each star. 
The internal uncertainty of the RV for each spectrum is estimated as the standard error of the mean, of the order of 0.1-0.3 km/s. 
The final RV for each target is obtained as the weighted mean of the values obtained from the two setups.
The accuracy of the wavelength calibration has been checked by measuring 
the position of the strong emission sky line at 6300.3 \AA\ in the HR13 setup, finding no significant offset.
No sky emission lines are available in the HR11 setup and we cannot directly check the accuracy of the wavelength 
calibration. However, the RVs obtained from the two setups agree each other  with an average difference between RV from 
HR11 and HR13 of +0.12$\pm$0.06 \kms ($\sigma$=~0.8 \kms ).
This excludes 
any offset for the two setups and confirming the accuracy also of the RVs from HR11 spectra.

\subsection{Chemical abundances}

The chemical abundances of Na, Mg (from the line at 5711 \AA), Si, Ca, Ti, Fe, Ni and Zr have been derived 
from the measure of the equivalent widths (EWs) of unblended lines by using the 
code {\tt GALA} \citep{m13g}. EWs have been measured by using the code {\tt DAOSPEC} \citep{stetson08}.

For species whose lines are affected by blending (O and the Mg lines at 6318-19 \AA\ ) 
or by hyperfine/isotopic splitting (Sc, V, Cu, Ba and La), abundances have been derived using our own code {\tt SALVADOR} 
that performs a $\chi^2$-minimisation between the observed lines and a grid of synthetic spectra calculated on-the-fly 
with the code {\tt SYNTHE} \citep{sbordone04} and including all the atomic and molecular lines available 
in the Kurucz/Castelli linelists. 
For all the species investigated, when the lines are not clearly detectable, 
we provide upper limits based on the comparison between observed and synthetic spectra.

\begin{figure}[h]
\includegraphics[width=\hsize,clip=true]{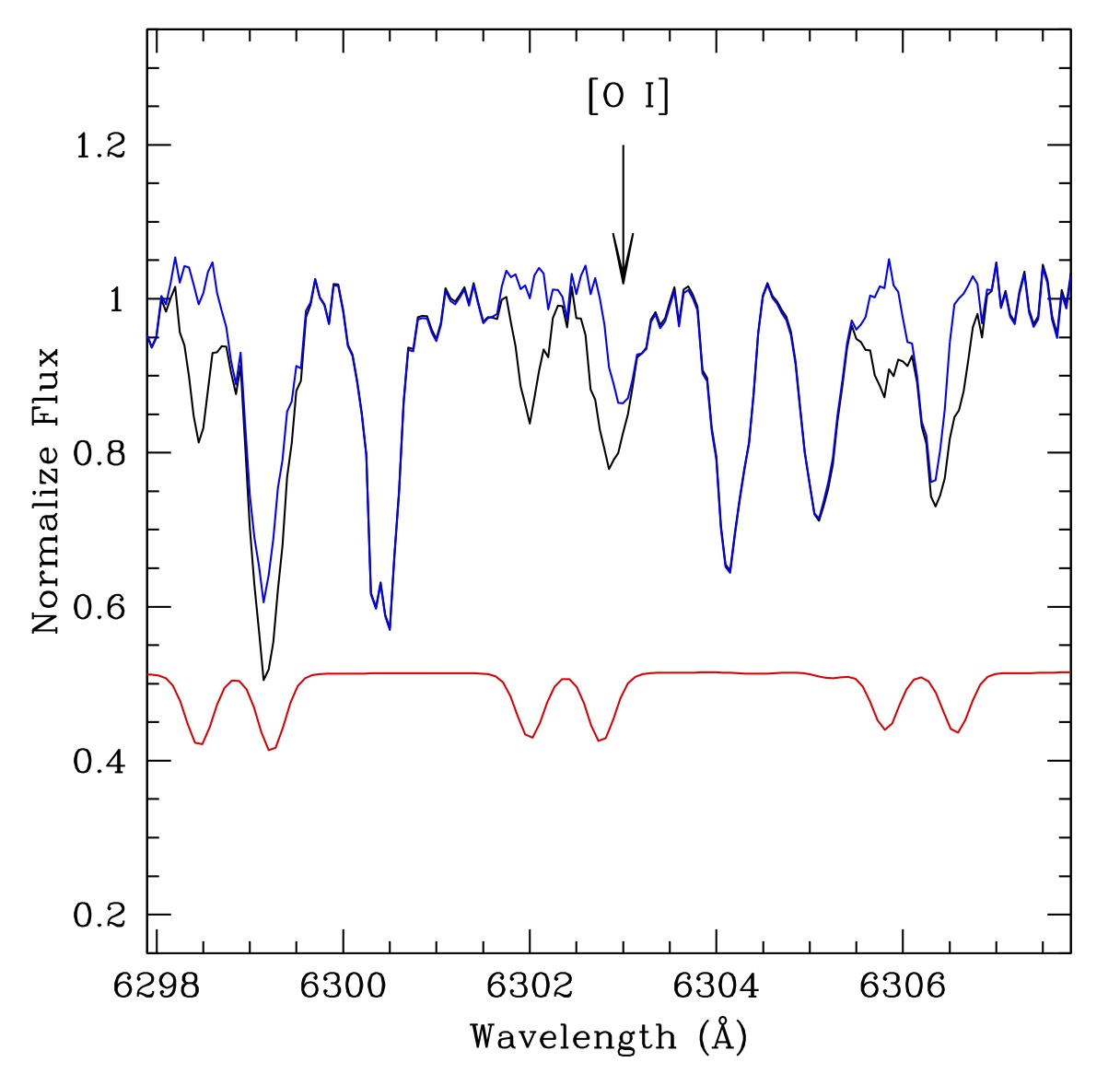}
\caption{Example of the telluric correction around the [O I] line in the star FLD-339\_466: 
black curve is the original (not corrected for RV) spectrum, the blue curve is the spectrum 
corrected for the telluric lines and the red one is the synthetic spectrum of the Earth atmosphere (shifted 
for sake of clarity).}
\label{tell}
\end{figure}

The [OI] line at 6300 \AA\ can be also contaminated by telluric lines, depending on the stellar 
RV. This possible contamination has been checked with suitable synthetic spectra of the Earth atmosphere 
calculated with the {\tt TAPAS} tool \citep{bertaux14}.
These synthetic spectra are calculated assuming the 
appropriate date of observation and airmass of our targets, 
in order to account for the proper weather conditions of the observations.
In case 
of contamination,  the line profile has been cleaned by dividing the observed spectrum by the telluric one and 
visually checking that no discontinuities were introduced. Fig.~\ref{tell} shows an example of a stellar spectrum 
around the [OI] line before and after the telluric correction.


In the calculation of the synthetic spectra used to measure the oxygen abundance, 
the Ni abundance of each star has been included to account for the blending of the O feature with a Ni line.

Mg abundances have been obtained for most of the stars from the EW of the line 
at 5711 \AA\ . As discussed in \citet{minelli21}, this transition is heavily saturated 
for giant stars with [Fe/H] $>$ --1.0 dex. For the latter stars, the Mg triplet at 6318-19 \AA\ 
should be preferred because these lines are still sensitive to the Mg abundance. 
Therefore, for the stars for which the Mg line at 5711 \AA\ turns out to be saturated, 
the Mg abundance has been derived from the Mg triplet using spectral synthesis 
in order to include the contribution of the close auto-ionization Ca  line.

Finally, only for the Na lines used here (5682-88 \AA\ and 6154-60 \AA\ ) 
we corrected the derived abundances for departures from the LTE assumption 
applying the corrections  by \citet{lind11}.

The abundances are referred to the solar ones, taking as reference the values from \cite{Grevesse1998}, 
apart from oxygen for which the adopted value is from \cite{Caffau2011}.

\subsection{Abundance uncertainties}

In the determination of the uncertainties in each derived abundance ratio 
we take into account two main sources of error, namely the errors arising 
from the measurement procedure (EW or spectral synthesis) and those arising 
from atmospheric parameters.

(1)~Uncertainties related to the measurement procedure are computed as 
the dispersion of the mean normalised to the root mean square of the number 
of used transitions. Properly, this term includes both uncertainties from line fitting 
and from adopted log~gf values.
For the elements measured from the EWs and for which only one line is available, 
the DAOSPEC uncertainty associated to the Gaussian fitting procedure 
(corresponding to 1$\sigma$ of the fit residuals) is assumed as internal error.
For the elements (O and La) for which only one transition has been measured using spectral synthesis, 
the internal error has been estimated by means of Monte Carlo simulations, creating a sample of 
500 synthetic spectra with a Poissonian noise that reproduces the observed SNR and repeating 
the line-fitting procedure. The dispersion of the abundance distribution obtained from these 
noisy synthetic spectra is assumed as 1$\sigma$ uncertainty.\\
(2)~Uncertainties due to atmospheric parameters have been estimated by repeating
the analysis by varying each time a given parameter of the corresponding 1$\sigma$ error 
and keeping fixed the other parameters. 

 These two sources of uncertainties have been 
added in quadrature. 
Since the abundance of the species X is expressed as abundance ratios [X/Fe], also 
the uncertainties in the Fe abundance have been taken into account. 
The final errors in [Fe/H] and [X/Fe] abundance ratios are calculated as follows:
\begin{equation}
  \sigma_{[Fe/H]} =
\sqrt{\frac{\sigma_{Fe}^2}{N_{Fe}} +  (\delta^{\rm T_{\rm eff}}_{Fe})^2 + (\delta^{log~g}_{Fe})^2 + (\delta^{\rm v_{\rm t}}_{Fe})^2}
\end{equation}

\begin{equation}
\begin{split}
&\sigma_{[X/Fe]}=\\
&\sqrt{\frac{\sigma_{X}^2}{N_{X}} + \frac{\sigma_{Fe}^2}{N_{Fe}} + (\delta^{\rm T_{\rm eff}}_X - \delta^{\rm T_{\rm eff}}_{Fe})^2 + (\delta^{log~g}_X - \delta^{log~g}_{Fe})^2 + (\delta^{\rm v_{\rm t}}_X - \delta^{\rm v_{\rm t}}_{Fe})^2}
\end{split}
\end{equation}

where $ \sigma_{X,Fe}$ is the dispersion around the mean of the chemical abundances, $N_{X,Fe}$ 
is the number of lines used to derive the  abundances and $\delta^{i}_{X,Fe}$ are the abundance 
variations  obtained modifying the atmospheric parameter {\sl i}.

\subsection{Abundances of the MW control sample}

Table~\ref{gcs} lists the average abundance ratios, together with the standard deviation 
and the average uncertainty in the abundance ratio, for the 5 GCs 
of the MW control sample.
We compared the atmospheric parameters and [Fe/H] of the analysed stars with those by \citet{carretta09} and 
\citet{carretta14} that analysed the same spectroscopic dataset.
The average differences between our analysis and the literature ones are +52$\pm$11 K ($\sigma$=~50 K) for \teff\ , 
--0.01$\pm$0.01 ($\sigma$=~0.03) for log~g, +0.07$\pm$0.04 \kms ($\sigma$=~0.19 \kms ) for \vt\  
and --0.03$\pm$0.02 dex ($\sigma$=~0.07 dex) for [Fe/H].

\begin{table*}[!h]
\caption{Average abundance ratios, the corresponding standard deviation and the average uncertainty for the 
five GCs of the MW control sample.}            
\label{gcs}     
\centering                 
\begin{tabular}{c   cc  cc cc cc  cc c}        
\hline\hline                 
\centering                  
{\rm Ratio}  & NGC~104  &  & NGC~1851  &  & NGC~5904 &  & NGC~1904 &  & NGC~4833 &  & \\
\hline   
   &  $< >$ & $\sigma$  &  $< >$ & $\sigma$  &  $< >$ & $\sigma$  &  $< >$ & $\sigma$  &  $< >$ & $\sigma$  & $<\sigma_{[X/Fe]}>$\\
\hline   
{\rm [Fe/H]}    &  --0.84    & 0.02   & --1.15   &  0.03  &  --1.29   &  0.01  &  --1.57   &  0.04  &   --2.11  &   0.03 & 0.07 \\
{\rm [O/Fe]}    &   +0.42    & 0.04   &  +0.40   &  0.03  &  +0.47   &  0.04  &   +0.54       &  0.04  &    +0.59  &   0.03  & 0.08 \\
{\rm [Na/Fe]}   &   +0.00    & 0.03   & --0.15   &  0.04  & --0.35   &  0.04  & --0.40   &  0.03  &  --0.50  &   0.05  & 0.08 \\
{\rm [Mg/Fe]}   &   +0.31    & 0.04   &  +0.34   &  0.04  &  +0.33   &  0.03  & +0.35   &  0.04  &  +0.38  &   0.02  & 0.10 \\
{\rm [Si/Fe]}   &   +0.28    & 0.03   &  +0.26   &  0.05  &  +0.28   &  0.02  & +0.30   &  0.06  &  +0.46  &   0.06  & 0.11 \\
{\rm [Ca/Fe]}   &   +0.29    & 0.04   &  +0.25   &  0.03  &  +0.26   &  0.03  & +0.25   &  0.01  &  +0.26  &   0.06  & 0.09\\
{\rm [Sc/Fe]}   &   +0.35    & 0.03   &  +0.17   &  0.04  &  +0.26   &  0.05  & +0.13   &  0.03  &  +0.28  &   0.04  & 0.08 \\
{\rm [Ti/Fe]}   &   +0.24    & 0.02   &  +0.06   &  0.01  &  +0.16   &  0.03  & +0.15   &  0.01  &  +0.20  &   0.06  & 0.09 \\
{\rm [V/Fe]}    &   +0.19    & 0.04   & --0.15   &  0.03  & --0.06   &  0.05  & --0.05   & 0.03  & --0.09  &   0.03   & 0.10\\
{\rm [Ni/Fe]}   &  --0.04    & 0.02   & --0.10   &  0.04  & --0.11   &  0.03  & --0.08   &  0.02  &  --0.11  &   0.04  & 0.06 \\
{\rm [Cu/Fe]}   &  --0.02    & 0.04   & --0.41   &  0.02  & --0.37   &  0.05  & --0.52  &  0.03  &  --0.60  &   0.04  & 0.08 \\
{\rm [Zr/Fe]}   &   +0.30    & 0.04   &  +0.11   &  0.03  &  +0.10   &  0.03  & +0.14   &  0.06  &   +0.06  &   0.06  & 0.12 \\
{\rm [Ba/Fe]}   &   +0.04    & 0.05   &  +0.13   &  0.04  &  +0.08   &  0.06  & +0.10   &  0.04  &   +0.31  &   0.05  & 0.12 \\
{\rm [La/Fe]}   &   +0.29    & 0.03   &  +0.35   &  0.05  &  +0.24   &  0.04  & +0.14   &  0.03  &   +0.27  &   0.06  & 0.08 \\
\hline                                   
\end{tabular}
\end{table*}

\section{RV and [Fe/H] distributions}

\subsection{RV distribution}

\begin{figure}[!h]
\includegraphics[width=\hsize,clip=true]{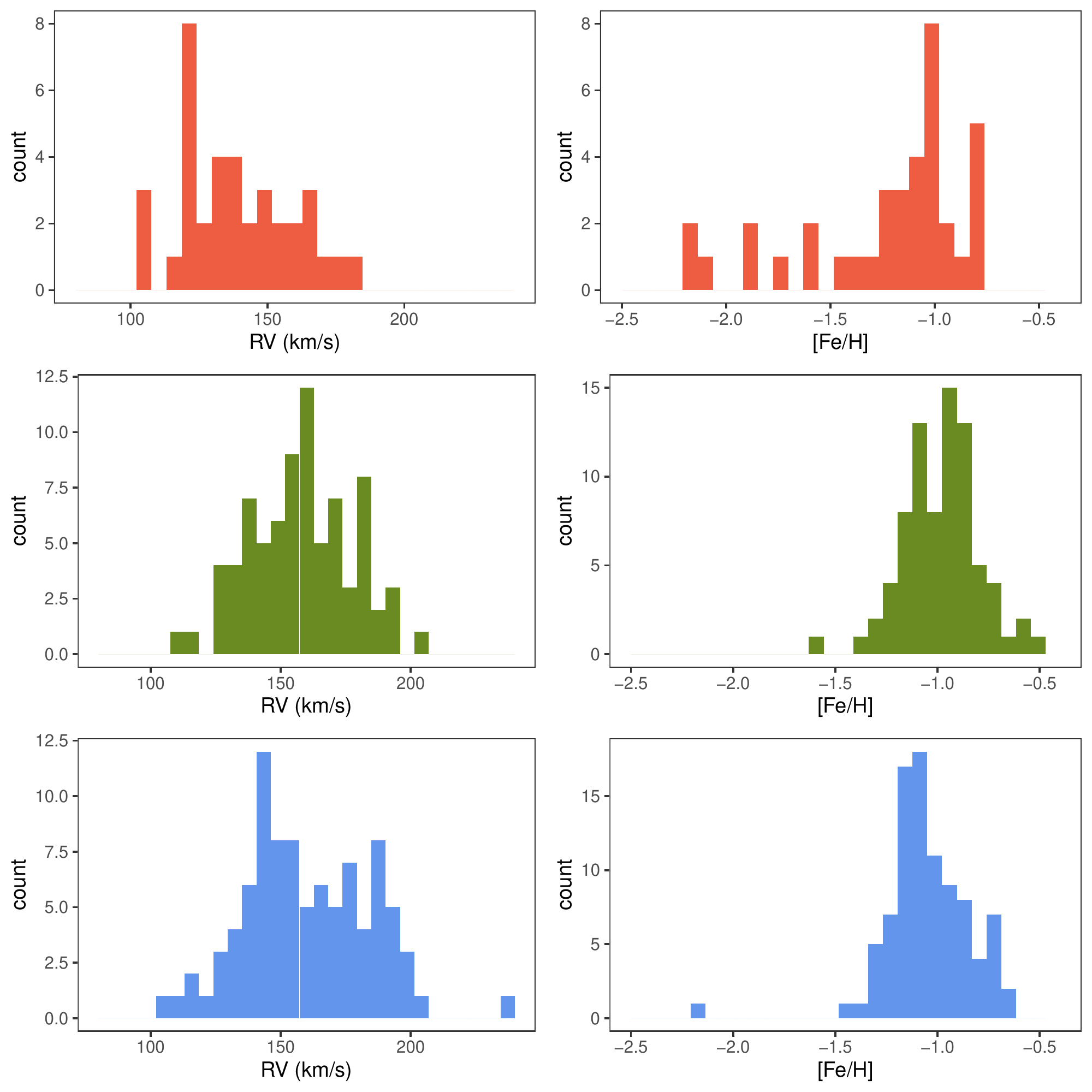}
\caption{RV and [Fe/H] distributions (left and right panel, respectively) of the three fields. 
Colours indicate the different fields: FLD-121 (red), FLD-339 (green) and FLD-419 (blue).}

\label{discr}
\end{figure}

\begin{figure*}[htbp]
\includegraphics[clip=true,width=0.8\textwidth]{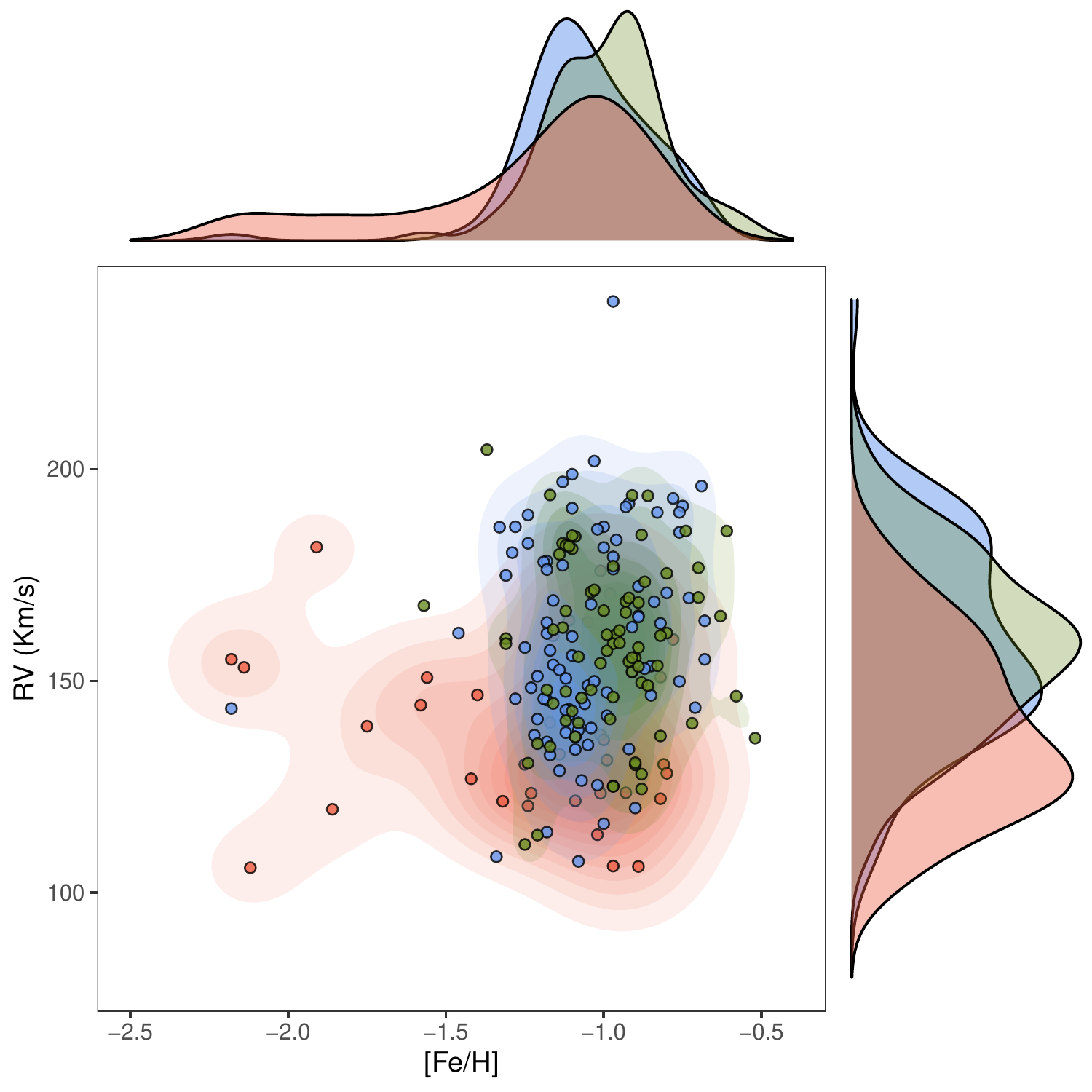}
\caption{RVs are plotted against [Fe/H] for target stars in the central panel of the figure. 
Colour-shaded areas denote the contours of the three clusters RV vs [Fe/H] distributions. 
Side plots show the kernel distributions of the RV (right-hand panel) and the [Fe/H] values (top panel) 
for each cluster. Same colours of Fig.~\ref{discr}}.
\label{rvfe}
\end{figure*}

According to previous spectroscopic studies  \citep{harris06,carrera08,dobbie14a,deleo20,hasselquist21} 
we identified as members of the SMC those stars with RV between +80 and +250 \kms . 
The membership is confirmed also by the proper motions measured from Gaia EDR3 \citep{brown21}.
We exclude from the chemical analysis stars members of the GC associated to each field 
(these stars will be discussed in a forthcoming paper of the series), stars with spectra contaminated 
by prominent TiO or $C_2$ molecular bands or with too low SNR. The final sample discussed in this work includes a total of 206 stars out 
of the 320 observed stars. The RV and [Fe/H] for this sample are listed in Table~\ref{stars}.  
Fig.~\ref{discr} and ~\ref{rvfe} shows the RV and [Fe/H] discrete and kernel density distributions of the three SMC fields. 
The advantage of the latter representation is that the distribution is 
independent of the choice of the bin width and of the starting bin, at variance with the discrete distributions.

The RV distributions of the three fields appear significantly different with each other, 
both in terms of the main peak and shape. The RV distribution of FLD-121 peaks at RV$\approx$+125 \kms\ , that of FLD-339 
displays a peak at RV$\approx$+160 \kms\ , while that of FLD-419 
exhibits two distinct peaks, the main one at $\approx$+150 \kms\ and the second one at $\approx$+180 \kms\ . 
A Kolmogorov-Smirnov test performed on these distributions confirms that the RV distributions of FLD-339 and FLD-419 
are significantly different with respect to that of FLD-121 (with statistic significance larger than 99.9\%), 
while we cannot reject the hypothesis that FLD-339 and FLD-419 may derive from the same population.

The differences in the peak of these three RV distributions 
are compatible with the rotation pattern of the SMC as inferred 
from low-resolution spectroscopic surveys of giant stars \citep{dobbie14a,deleo20}, 
from the HI column density map \citep{diteodoro19} and from the APOGEE results from 17th Data Release of the Sloan Digital Sky Survey 
\citep{sdss22}.  All these studies show that the western side of the SMC, where FLD-121 is located, has a lower 
velocity with respect to the eastern side. However, the presence of multiple peaks, clearly visible in the distribution 
of FLD-419, seems to suggest a more complex kinematic pattern (as discussed below).

\subsection{[Fe/H] distribution}
The [Fe/H] distribution of the entire sample is peaked at [Fe/H]$\sim$--1.0 dex, 
with about 95\% of the stars having [Fe/H] between --1.5 and --0.5 dex 
and with a weak but extended metal-poor tail 
reaching [Fe/H]$\sim$--2.2 dex. This distribution is qualitatively similar to those 
obtained from low-resolution spectra using Ca~II triplet \citep{carrera08,dobbie14a,parisi16} and that from APOGEE data \citep{nidever20}. 
However, similar to what we see with the RV distributions, when the individual fields are considered, the metallicity distributions 
appear different with each other. 
The distributions of FLD-339 and FLD-419 are confined between --1.5 and --0.5 dex, 
with only one star per field ($\sim$1\%) with [Fe/H]$<$--1.5 dex . 
On the other hand the distribution of FLD-121 
ranges from --0.8 dex down to --2.2 dex, with $\sim$20\% of the stars more metal-poor than --1.5 dex. Note that the APOGEE field 47Tuc, 
superimposed to our field FLD-121, exhibits a lower fraction of metal-poor stars, $\sim$2\%, 
probably reflecting some selection bias against metal-poor stars in the APOGEE observations.\\
The peaks of the distributions of FLD-339 and FLD-419 are separated by $\sim$0.2 dex and located at [Fe/H]$\sim$--0.9 and $\sim$--1.1 dex,
respectively. Also, the two distributions seem to be not symmetric, with the presence a secondary peak at [Fe/H]$\sim$--1.1 dex 
in FLD-339 
and a heavily-populated metal-rich tail or a secondary peak in FLD-419 (see Sec.~\ref{subst}).

\subsection{[Fe/H] distribution and the age-metallicity relation}

We try to interpret the derived [Fe/H] distributions in terms of ages, using as a guidance the SF histories recovered 
from Hubble Space Telescope  \citep{noel07,sabbi09,cignoni12,cignoni13} and ground-based \citep{massana22} photometry, 
and the theoretical age-metallicity relations available for the SMC \citep{Pagel1998,tb09,cignoni13}.

All these works agree that the early epochs of the SMC have been characterised by a significant SF activity followed by 
a long quiescent period, interrupted between $\sim$3 and $\sim$4 Gyr ago by significant SF episodes, 
likely due to some merger events.
The oldest SMC GC, NGC~121, has an age of $\sim$10.5$\pm$0.5 Gyr \citep{glatt08} and 
a metallicity of [Fe/H]$\sim$--1.2/--1.3 dex \citep[][A. Minelli et al. in prep.]{dalex16}.
This suggests that the SF activity in the first Gyrs was able to increase the metallicity to values as high 
as [Fe/H]$\sim$--1.2/--1.3 dex. 
We can consider the SMC field stars in our sample with [Fe/H]$<$--1.3 dex (which are almost all located in FLD-121) 
as formed in the first 1-2 Gyr of the life of the galaxy. 

\begin{figure}[!h]
\includegraphics[width=\hsize,clip=true]{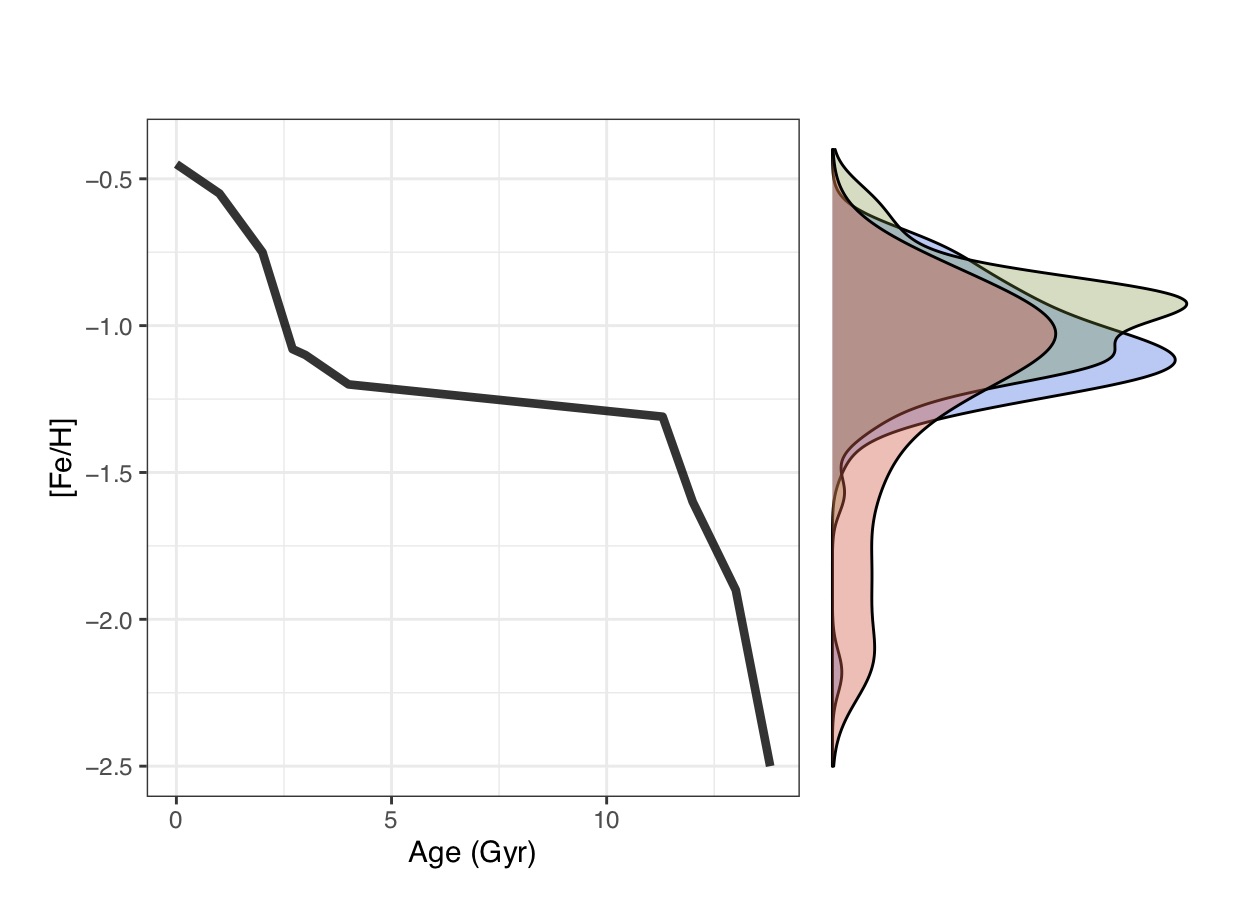}
\caption{Main panel: age-metallicity relation by \citet{Pagel1998}. 
Side panel: the kernel [Fe/H] distributions for the 
individual SMC field stars discussed in this work.}
\label{amr}
\end{figure}

The subsequent evolution of the SMC and the corresponding metallicity distribution 
can be interpreted in the light of the theoretical age-metallicity relations: 
Fig.~\ref{amr} shows that by \citet{Pagel1998} assuming a burst of SF at an age of $\sim$4 Gyr.
After a long period characterised by a low SF efficiency (and where the metallicity remains almost constant), 
the SF in the SMC re-ignites with a prominent burst,
likely triggered by the first close encounter between SMC and LMC \citep{bekki04,bekki05}. 
The most recent SF history for the SMC provided by \citet{massana22} using the SMASH photometry identified the re-ignition 
of the SF at $\sim$3.5 Gyr ago, simultaneously in both the Clouds.
The stars with [Fe/H]$>$--1.3 dex analysed here should be a mixture of stars with different ages 
(from $\sim$1 to $\sim$10-11 Gyr). It is not easy to separate the different populations in terms 
of age due to the almost constant [Fe/H] over a large age range. 
\citet{massana22} identified in the SF history of the SMC five peaks 
(at $\sim$3, 2, 1, 0.45 Gyr ago and one still ongoing) occurring simultaneously also
in the LMC. 
A fascinating possibility is that the different peaks in the 
metallicity distributions of FLD-339 and FLD-419 could be associated to some of these different bursts of SF. 
Finally, we can suppose that the stars with [Fe/H] around --0.6/--0.5 dex are 
likely formed with the burst at 1 Gyr. This is confirmed also by the metallicities of the stellar clusters 
with ages around 1 Gyr \citep[see e.g.][]{parisi22}

\subsection{Run of [Fe/H] with the distance}

\begin{figure}[htbp]
\includegraphics[width=\hsize,clip=true]{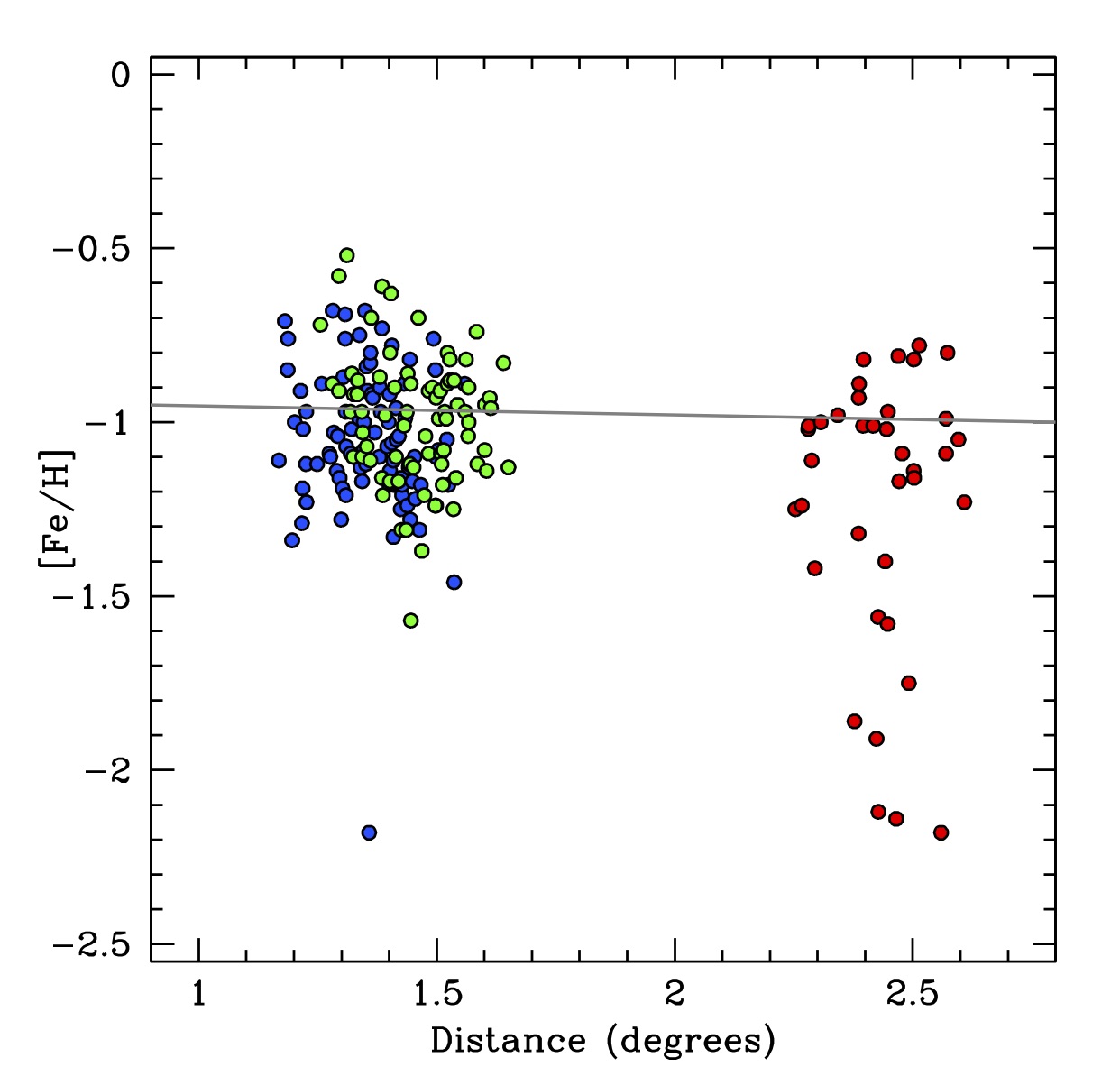}
\caption{Behaviour of [Fe/H] as a function of the projected distance from the SMC centre \citep{ripepi17}, 
same colours of Fig.~\ref{map}. The thick grey line is the linear fit for the metallicity gradient estimated by \citet{chou2020}.}
\label{grad}
\end{figure}

Previous spectroscopic studies \citep{carrera08,dobbie14a,parisi16,chou2020,grady21} found 
evidence of a shallow (from --0.03 to --0.07 dex/deg) metallicity gradient, within 3$^{\circ}$--5$^{\circ}$.
Fig.~\ref{grad} shows the run of [Fe/H] of the spectroscopic targets with their projected distance from 
the SMC centre \citep{ripepi17}. 
The mean metallicity in three fields is consistent with the shallow gradient previously proposed by \citet{chou2020}.
However two main differences between the external field FLD-121 
and the two internal ones are evident. First, in FLD-121 the fraction of metal-poor stars ([Fe/H]$<$--1.5 dex) 
is about $\sim$20\%, against $\sim$1\% in the other two fields. The fraction of metal-poor stars 
increases outward reflecting a larger fraction of old stars with respect to those formed subsequently 
during the long quiescent period and the recent SF bursts and that are preferentially 
confined in the innermost region of the SMC
\citep[see e.g.][]{rubele18}.

Second, in the metallicity distribution of FLD-121 there is a clear lack of stars with [Fe/H] between --0.8 
and --0.5 dex, instead detected in FLD-339 and FLD-419. Following the discussion above, these stars should 
have $\sim$1 Gyr (the youngest stars among the intermediate-age SMC populations).
Again, this is consistent with a scenario where the younger, metal-richer populations are progressively 
more concentrated toward the innermost regions. 
Age-metallicity gradients of this kind are quite common in dwarf galaxies \citep[see e.g.][and references therein]{taibi22}.

\subsection{Possible kinematic/chemically distinct sub-structures?}
\label{subst}

The distribution of the SMC stars in the RV-[Fe/H] plane seems to suggest the presence of sub-structures, 
in particular the two different peaks of the [Fe/H] distribution of FLD-339, the large and asymmetric [Fe/H] 
distribution of FLD-419 and the double-peak of the RV distribution of FLD-419.

We used the gaussian mixture package Mclust \citep{scrucca16}, within the R environment, to analyse the distribution 
of FLD-339 and FLD-419 stars in the [Fe/H] - RV space. 
Mclust choose the best model, both in terms of number and form 
\citep[e.g.,  equal or variable variance, orientation etc., see][]{scrucca16}
of the gaussian components, by means of the Bayesian Information Criterion.
Since we are interested in substructures within the bulk of the metallicity distribution we exclude from 
the analysis the two metal-poor outliers, one per field. 
While for FLD-339 a single elliptical gaussian model is the preferred solution, the [Fe/H] - RV distribution 
of FLD-419 is best described with two elliptical gaussian components with the same variance  both in [Fe/H] and RV. 
The gain of  this model with respect to a single elliptical gaussian is only marginal, in practice they provide 
an equally good representation of the data. Still, the solution synthesise  the properties of the hypothesised 
two components. The first component has ($\mu_{\mathrm{[Fe/H]}}$, $\mu_{\mathrm{RV}}$)= ( -0.85 dex, 171.8 \kms), 
and it accounts for 33\% of the sample, the second component has ($\mu_{\mathrm{[Fe/H]}}$, 
$\mu_{\mathrm{RV}}$)= ( -1.13 dex, 154.5 \kms ), 
accounting for the remaining 67\% of the sample. 
The standard deviations are $\sigma_{\mathrm{[Fe/H]}}$=~0.10 dex, and $\sigma_{\mathrm{RV}}$=~22.7 \kms . 
It seems that the most metal rich component has a larger systemic RV than its metal-poor counterpart.

As additional check, we performed a Kolmogorov-Smirnov test 
on the RV sub-populations of FLD-419, separated according to the metallicity of their member stars 
(and assuming [Fe/H]=--1.05 dex as a boundary between the two groups of stars). 
We obtained that the two RV distributions cannot be extracted from the same population with a significance of 98\%.

The size of the FLD-419 sample is not sufficient to put this odd result on sound statistical bases, 
still it may be suggestive of the presence of some chemo-kinematic substructures in the SMC along this line of sight.
In this respect, it is worth recalling that the SMC has a substantial 
line-of-sight depth, depending on the used tracers and ranging from a few kpc 
up to about 20 kpc \citep[see e.g.][]{degris15,sub17}.
Therefore, when we observe stars in an individual SMC field 
we are likely crossing different depths and we are sampling different 
populations in terms of kinematics and metallicity.


\section{Chemical abundance ratios}
\label{res}

We derived abundances of Na, O, Mg, Si, Ca, Sc, Ti, V, Fe, Ni, Cu, Zr, Ba and La for 206 SMC RGB stars. 
All the abundances, with the corresponding uncertainties, are available in the 
electronic form (Table 5). With respect to the APOGEE sample by \citet{hasselquist21} we measured a larger number 
of species, in particular Na, Sc, Ti, V, Cu, Zr, Ba and La, not included in that study.
Fig.~\ref{alfa1}-\ref{slow1} show the behaviour of derived abundance ratios as a function of [Fe/H] 
for the analysed SMC stars, highlighting stars belonging to the different  fields. These abundance ratios 
are compared with those obtained for the control sample of 5 Galactic GCs, adopting the same assumptions 
in the chemical analysis and therefore removing  most of the systematics of the analyses.  
This comparison allows us to highlight the real difference between SMC and MW stars of similar [Fe/H].
Additionally, we show abundance ratios for Galactic field stars from the
literature as reference.
The comparison with the literature is affected by the systematics among the different analyses 
(in terms of model atmospheres, solar abundance values, NLTE corrections, linelists, use of 
dwarf and giant stars). However, it is useful to display the overall trends in the MW based on a large number of stars.
In the following, we refer to the MW control sample to quantify the main 
differences and similarities between MW and SMC stars.

\subsection{Na}
\begin{figure*}
\centering
\includegraphics[scale=0.225]{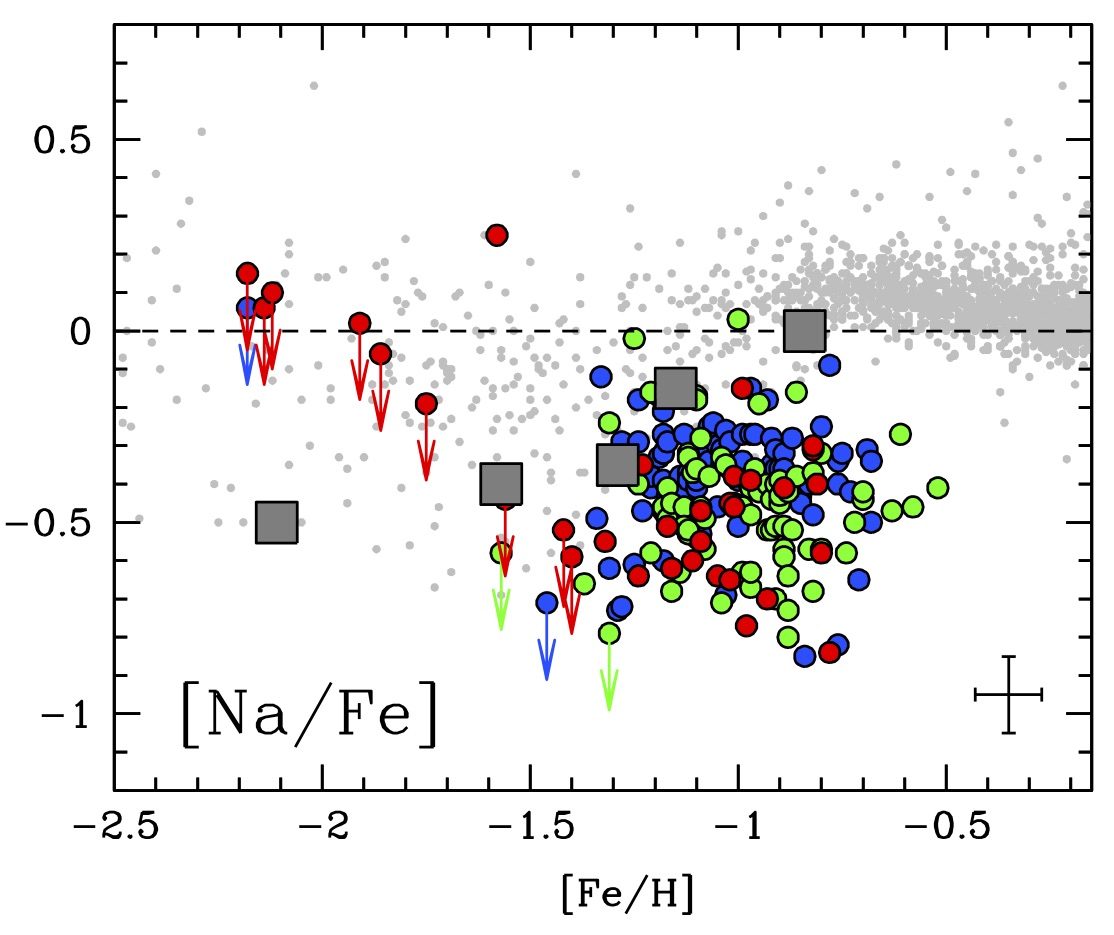}
\includegraphics[scale=0.225]{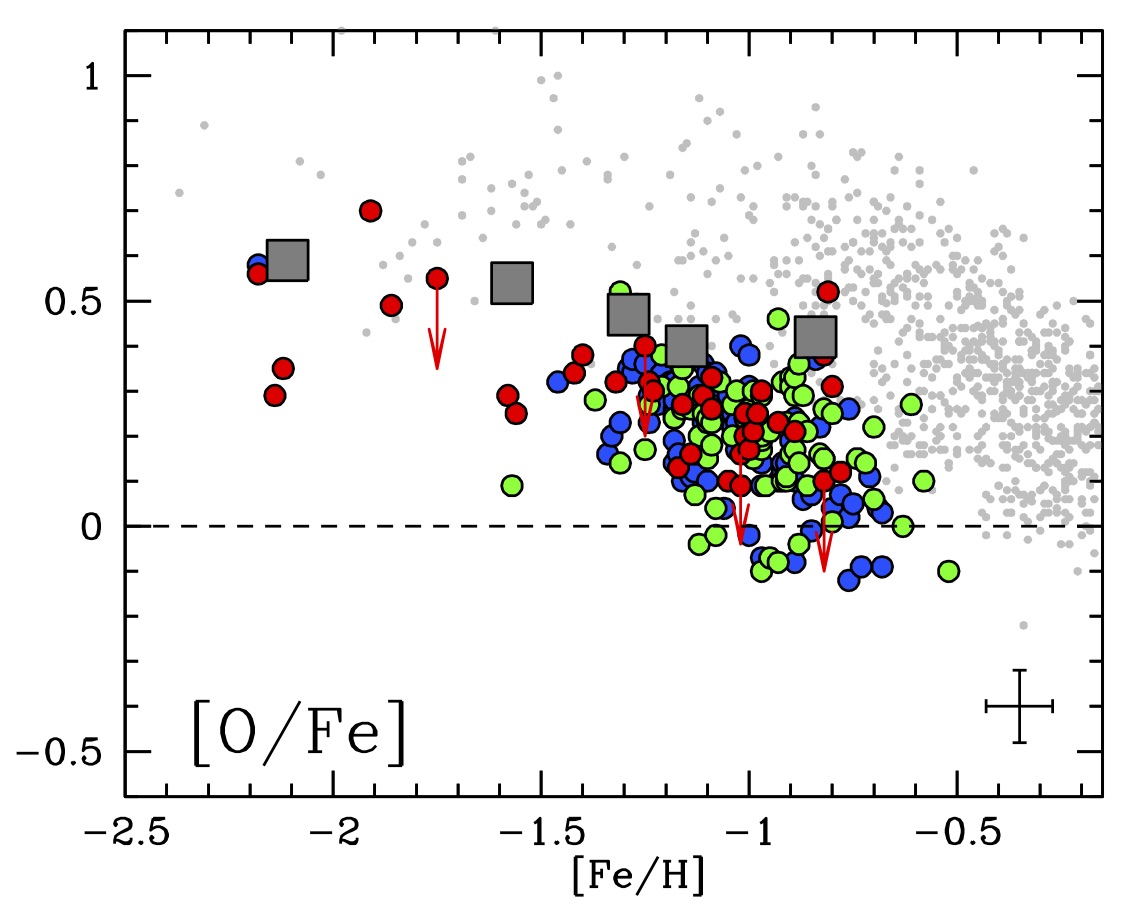}\\
\includegraphics[scale=0.225]{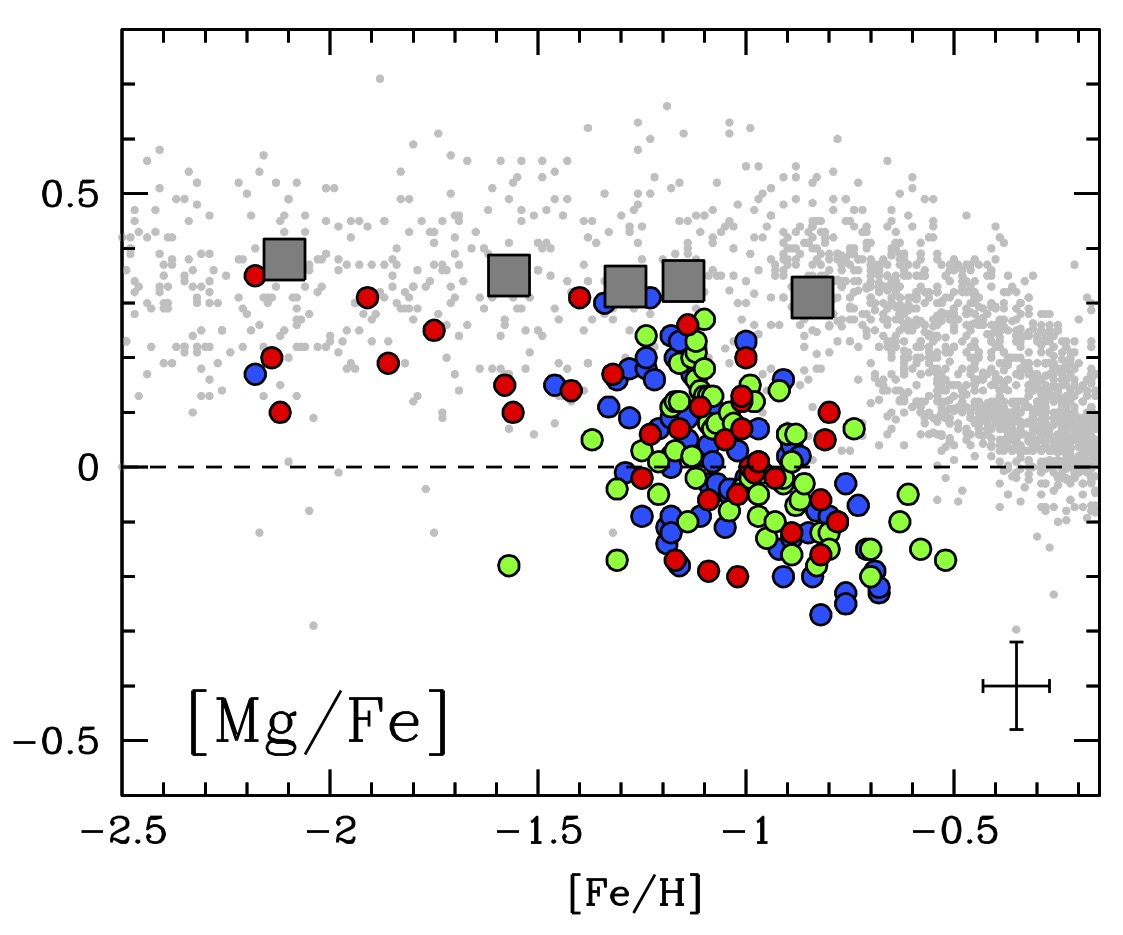}
\includegraphics[scale=0.225]{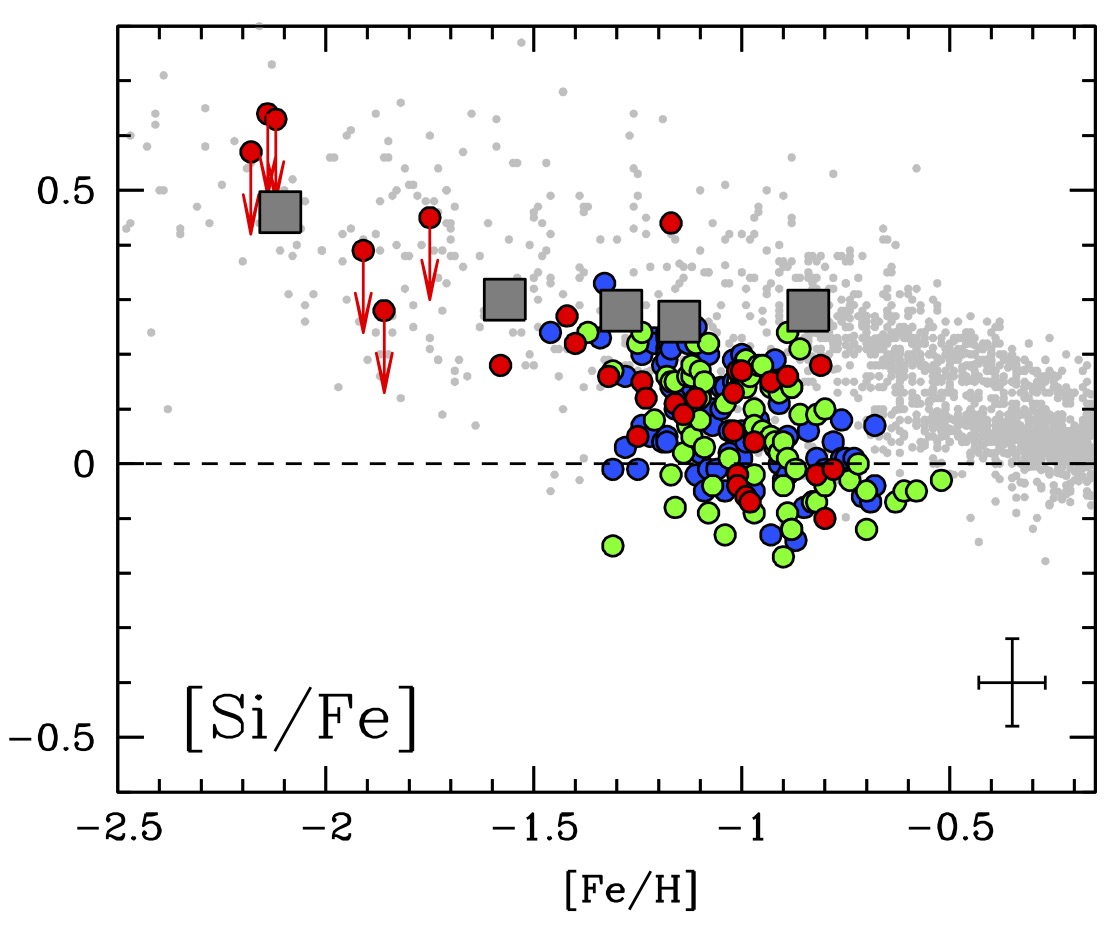}\\
\includegraphics[scale=0.225]{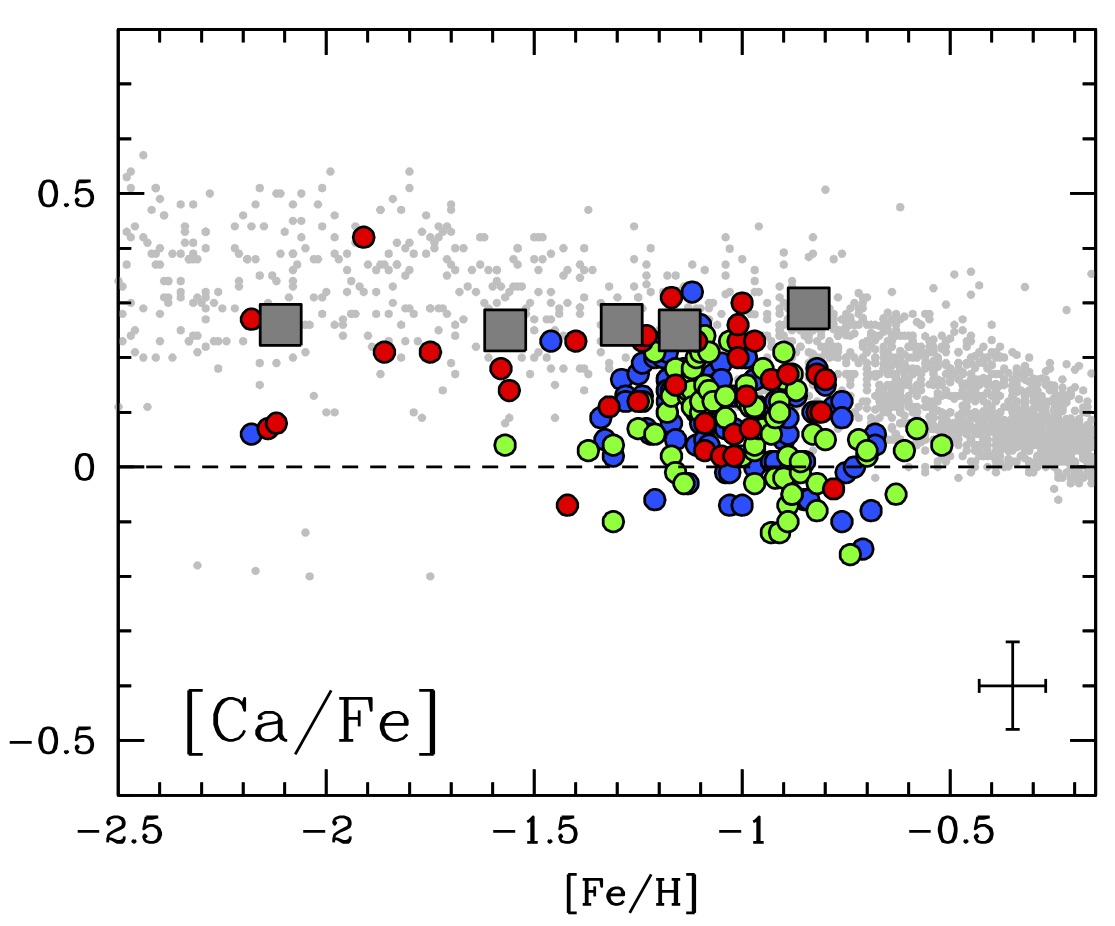}
\includegraphics[scale=0.225]{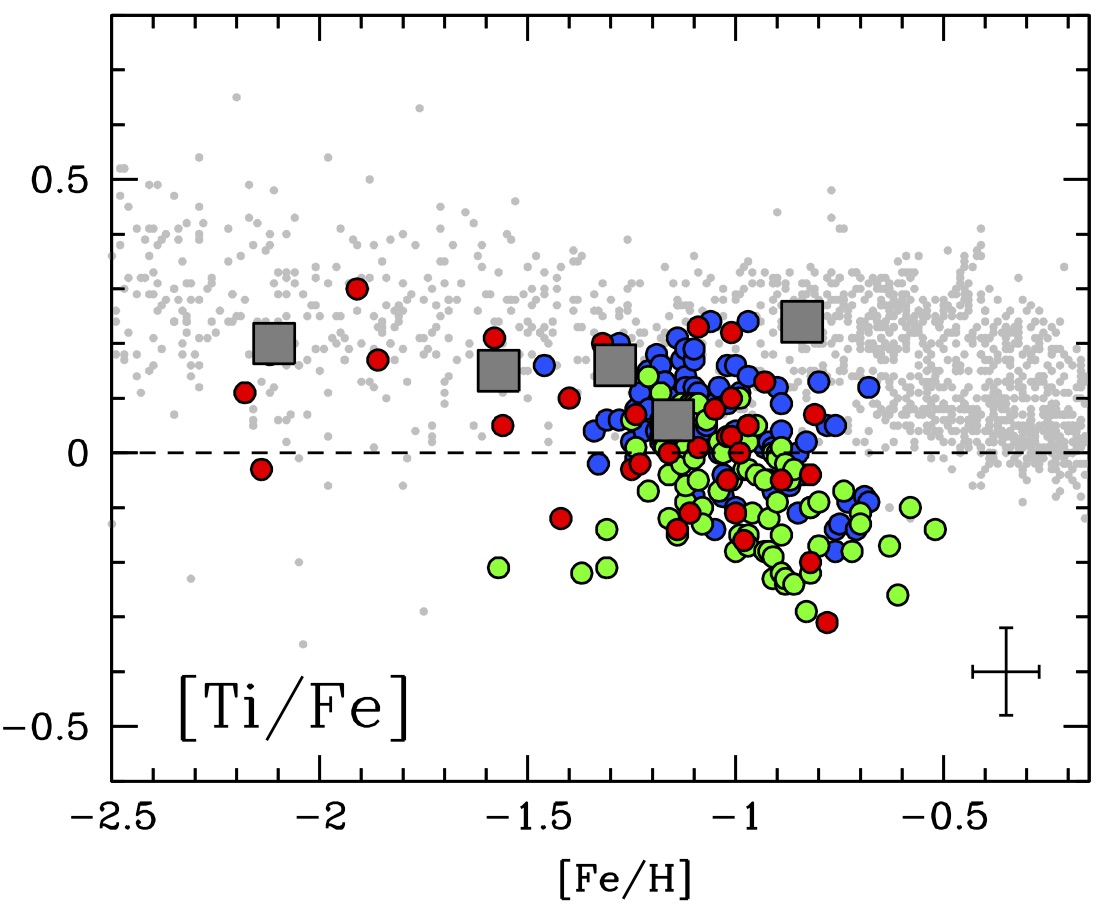}\\
\caption{Behaviour of the light element [Na/Fe] and $\alpha$-elements [O/Fe], [Mg/Fe], [Si/Fe], [Ca/Fe] and [Ti/Fe] abundance ratios 
as a function of [Fe/H] for SMC stars located in the fields FLD-419, FLD-339 and FLD-121 (blue, green and red circles, respectively). 
Arrows indicate upper limits. The errorbars in the bottom-right corner indicate the 
typical uncertainties. Grey squares are the average values for the five Galactic GCs of the control sample. 
Abundances of Galactic stars from the literature are also plotted as a reference: 
\cite{Edvardsson1993, Gratton2003, Reddy2003, Reddy2006, Bensby2005, Bensby2014} for all the elements, 
\cite{Fulbright2000, Stephens2002, Roederer2014} for Na, Mg, Si, Ca and Ti, 
\cite{Adibekyan2012} for Na, Mg, Si and Ca, \cite{Barklem2005} for Mg.}
\label{alfa1}
\end{figure*}

Sodium is mainly produced in massive stars during the hydrostatic C and Ne burning, with a strong 
dependence of its yields on the metallicity. Also, a smaller contribution is provided by asymptotic giant branch 
(AGB) stars. 
In Galactic stars (both in the control sample and in literature data), [Na/Fe] increases by increasing 
[Fe/H] until it reaches solar values around [Fe/H]$>$--1 dex. An offset is evident between the values in the control 
sample and in the literature, especially in the metal-poor regime and likely due to the different NLTE corrections.
Top-left panel of Fig.~\ref{alfa1} shows the distribution of [Na/Fe] of the observed targets. 
The bulk of the SMC stars exhibits sub-solar [Na/Fe] abundance ratios at any metallicities, with an average 
value of about --0.4/--0.5 dex, similar to the typical [Na/Fe] measured in the LMC stars \citep{vds13,minelli21} 
but at higher [Fe/H]. 
The low [Na/Fe] values measured in the SMC stars may point to a lower contribution by massive stars, besides 
the larger impact of Type Ia supernovae (SNe Ia) at low metallicities in dwarf galaxies \citep{tolstoy09}.
We observe a large scatter of [Na/Fe], not fully explainable within the typical uncertainties, and 
already detected in spectroscopic samples of LMC and SMC metal-rich stars 
\citep{pompeia08,vds13,minelli21,hasselquist21}. This scatter  could reflect that multiple sites of Na production are taking place.
Finally, we note a systematic difference between the median [Na/Fe] values in FLD-339 and FLD-419, where the latter displays 
[Na/Fe] higher by 0.1-0.15 dex. A systematic difference in [Na/Fe] of different regions of the parent galaxy has been also
observed in the LMC \citep{vds13}
with the stars in the LMC bar more enriched in [Na/Fe] by 0.2 dex with respect to the LMC disc stars.

\subsection{$\alpha$-elements}
The $\alpha$-elements are produced mainly in short-lived massive stars exploding as core-collapse supernovae (CC-SNe), 
while a minor fraction (depending on the element) is synthesised in SNe Ia.
Due to the time delay between the enrichment of the two classes of SNe,
the [$\alpha$/Fe] abundance ratios are the classical 
tracers of the relative timescales of the different SNe. In particular, the metallicity of the {\sl knee} 
(marking the onset of a significant chemical contribution by SNe Ia) can be used as a proxy of the SF efficiency of the galaxy 
\citep{tinsley79,matteucci86}.

\begin{figure}[h]
\includegraphics[scale=0.225]{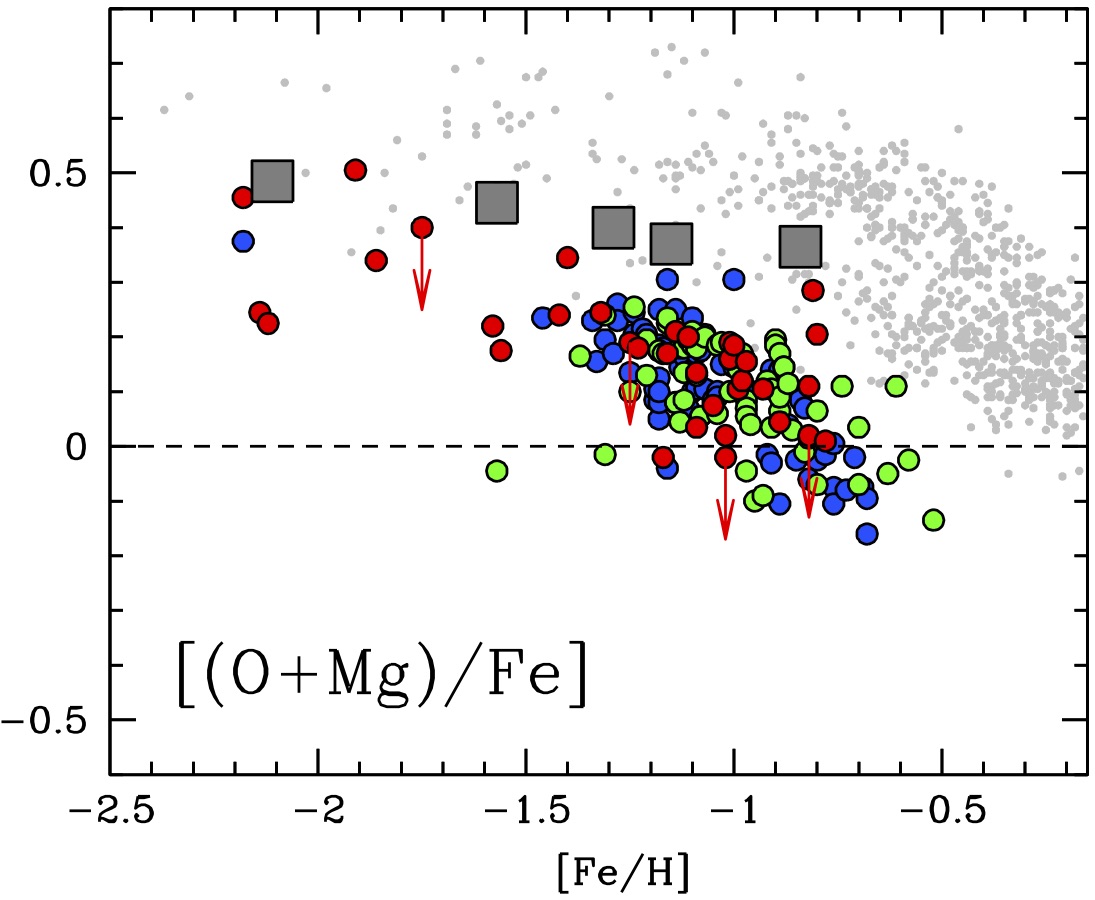}
\includegraphics[scale=0.225]{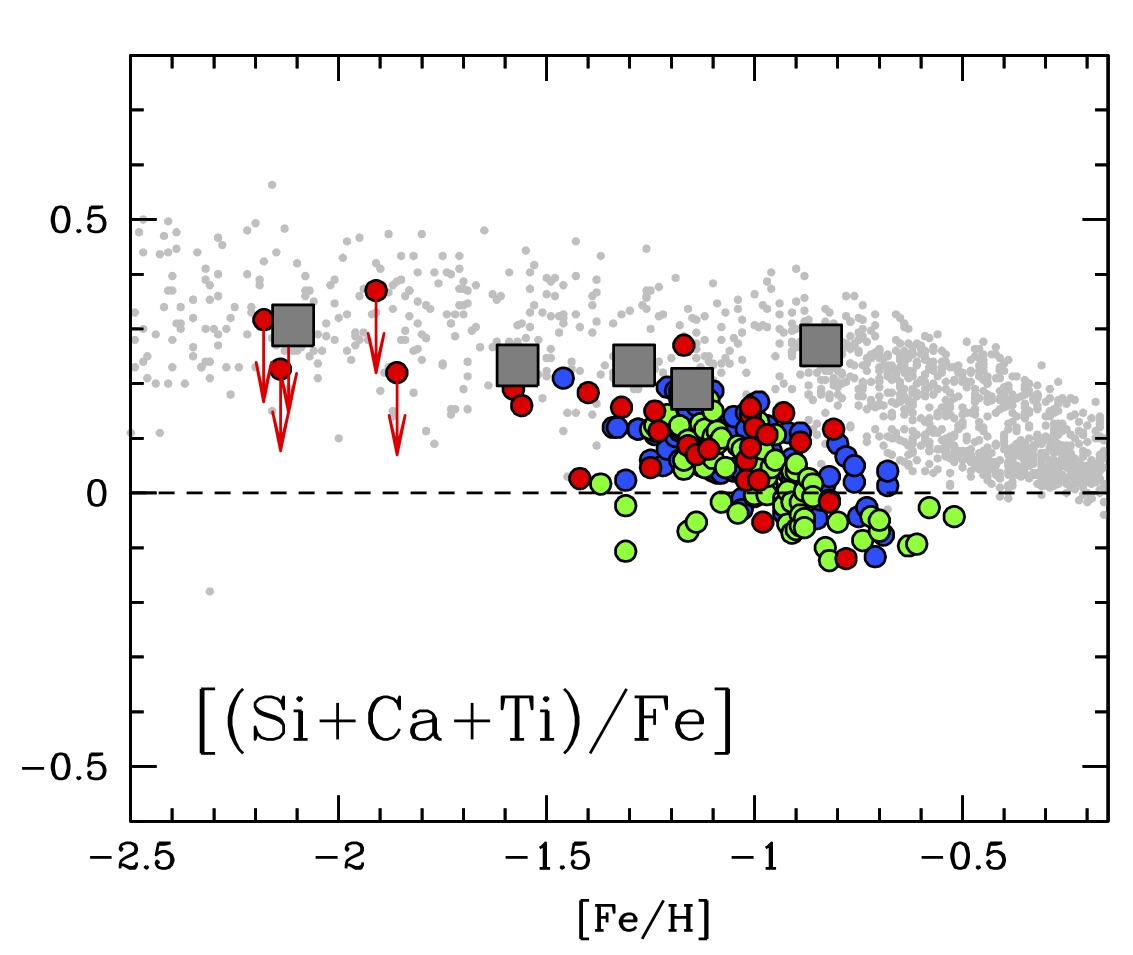}\\
\caption{Behaviour of the hydrostatic and explosive average 
[$\alpha$/Fe] abundance ratios 
as a function of [Fe/H]. 
Same symbols of Fig.~\ref{alfa1}.}
\label{alfat}
\end{figure}

O and Mg (the so-called hydrostatic $\alpha$-elements) are produced mainly in stars with masses larger than  $\sim$30-35 $M_{\odot}$ 
and without contribution by SNe Ia. On the other hand, Si, Ca and Ti (explosive $\alpha$-elements) are produced in less massive stars 
($\sim$15-25 $M_{\odot}$) and with a smaller (but not negligible) contribution by SNe Ia \citep[see e.g.][]{kobayashi20b}. 
Fig.~\ref{alfa1} shows the behaviour with [Fe/H] of individual [$\alpha$/Fe] abundance ratios, while 
Fig.~\ref{alfat} shows the run of the average values of hydrostatic and explosive [$\alpha$/Fe].
These abundance ratios in the SMC stars
clearly display a decrease by increasing the metallicity, moving from enhanced values for the most metal-poor stars ([Fe/H]$<$--1.5 dex) 
down to solar-scaled values in the dominant population. This trend is in contrast with that 
obtained by the APOGEE survey, where "there is a slight increase in [Mg/Fe] beginning at [Fe/H]$\sim$--1.3 dex, 
with a peak at [Fe/H]$\sim$--1.0 dex, followed by a slight decrease. The [O/Fe], [Si/Fe] and [Ca/Fe] abundance patterns 
are flat over this range" \citep{hasselquist21}.

The most metal-poor stars in our sample exhibit enhanced values of [$\alpha$/Fe]  
and in agreement with the results by \citet{nidever20} and \citet{reggiani21} for SMC stars of similar metallicity. 
Oxygen and magnesium, that are mainly produced by stars with masses 
larger than $\sim$30 $M_{\odot}$, are, however, slightly underabundant at low [Fe/H] with respect to the MW sample, 
which points to a lower contribution from the most massive stars to the overall chemical enrichment of the SMC.
The subsequent decrease of [$\alpha$/Fe] at higher [Fe/H] indicates  that these stars formed from a gas enriched by SNe Ia.
For stars with [Fe/H]$>$--1.5 dex the difference 
in [$\alpha$/Fe] between SMC and MW stars becomes more significant. In particular, the SMC-MW difference
is more pronounced for hydrostatic $\alpha$-elements, 
again suggesting a lower contribution  by stars with masses larger than 30-35 $M_{\odot}$ to the chemical enrichment of the SMC. 

We note, as for Na, that the metal-rich stars in FLD-419 are slightly enhanced in [Ti/Fe], by $\sim$0.1 dex, 
with respect to the stars of the other two fields with similar [Fe/H].

\subsection{Iron-peak elements}
Iron-peak elements are produced mainly in massive stars, through 
different nucleosynthesis paths \citep{limongi03,romano10,kobayashi20b}, and ejected 
in the interstellar medium both from normal CC-SNe and hypernovae. 
These elements are also partly produced by SNe~Ia on longer time scales 
\citep{leung18,lach20}.

\begin{figure*}[h]
\includegraphics[scale=0.225]{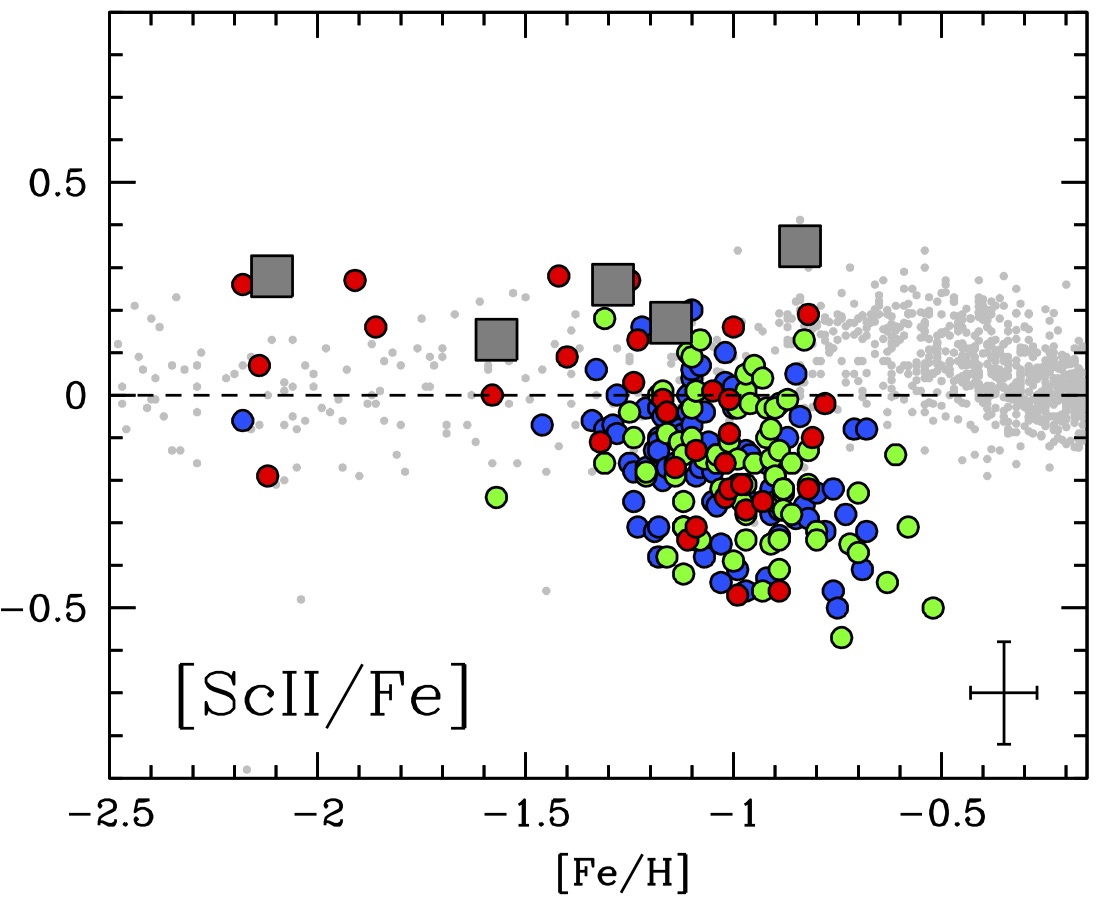}
\includegraphics[scale=0.225]{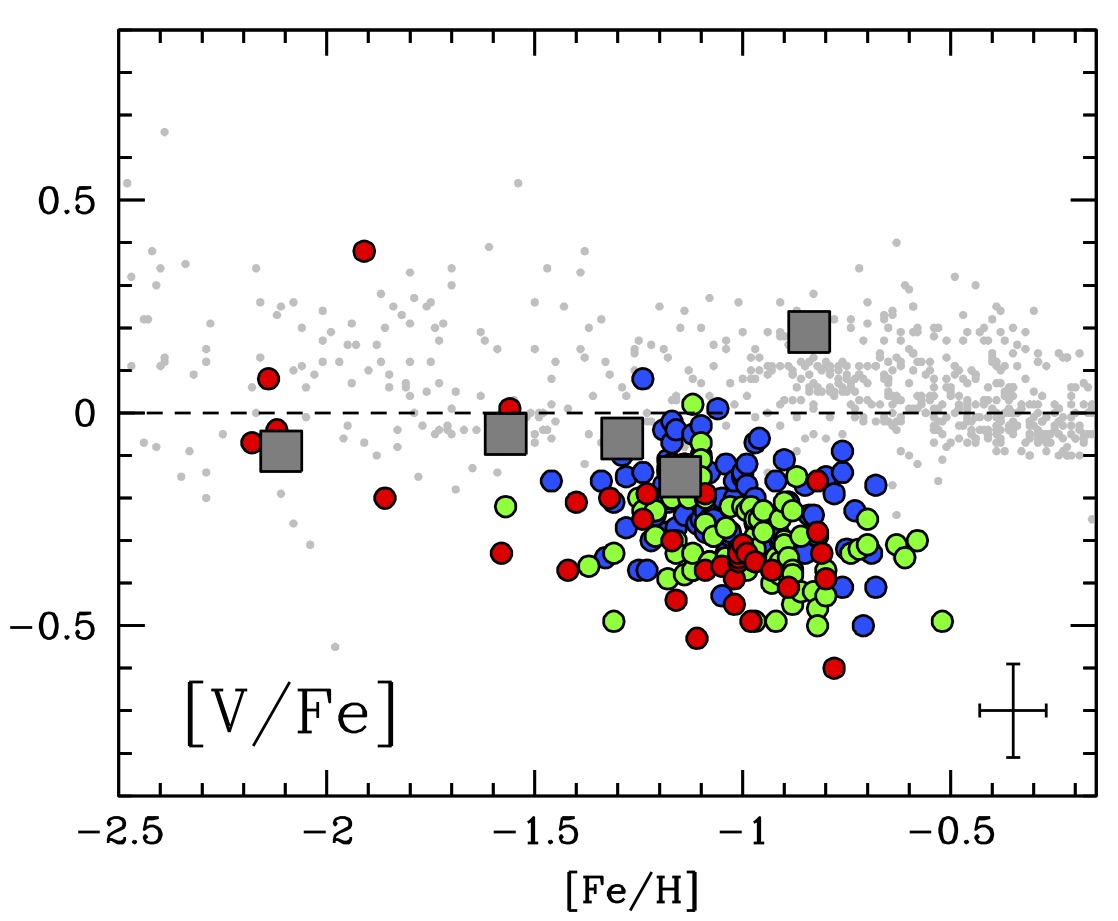}\\
\includegraphics[scale=0.225]{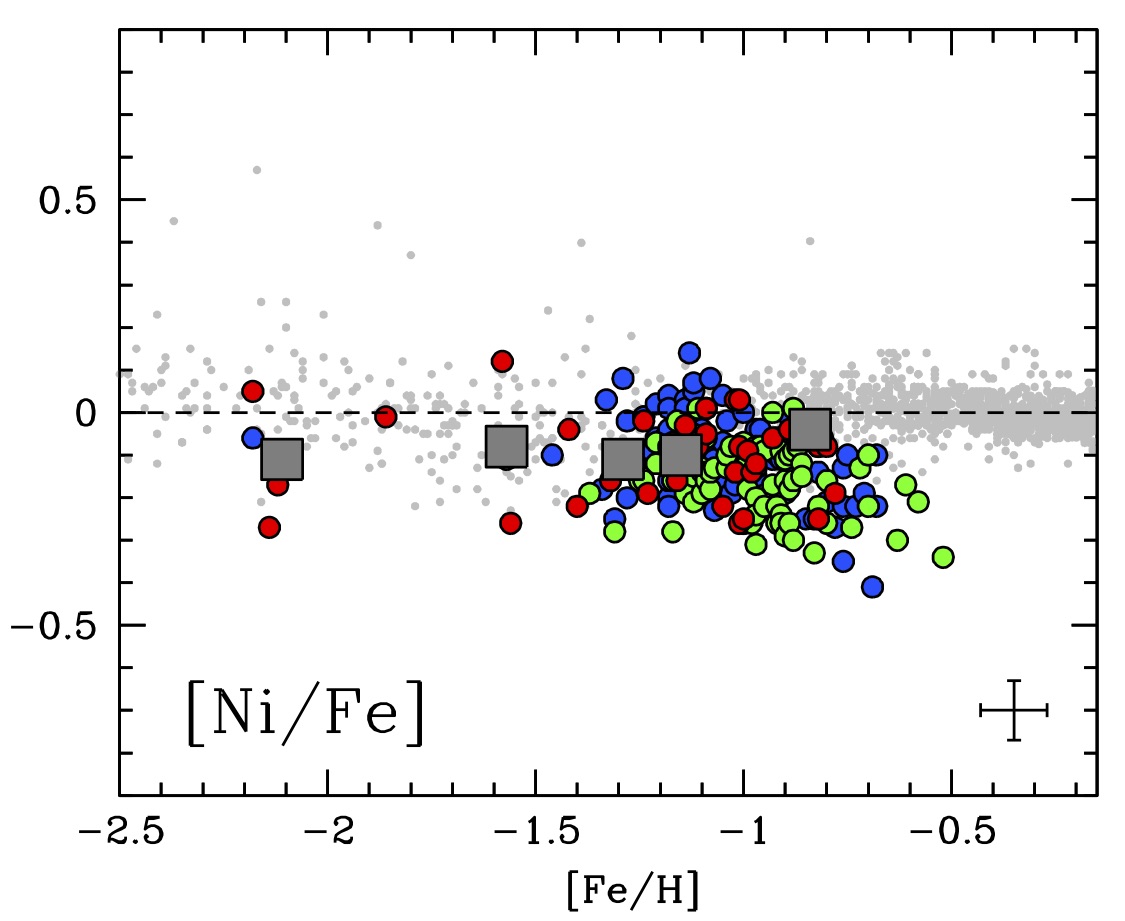}
\includegraphics[scale=0.225]{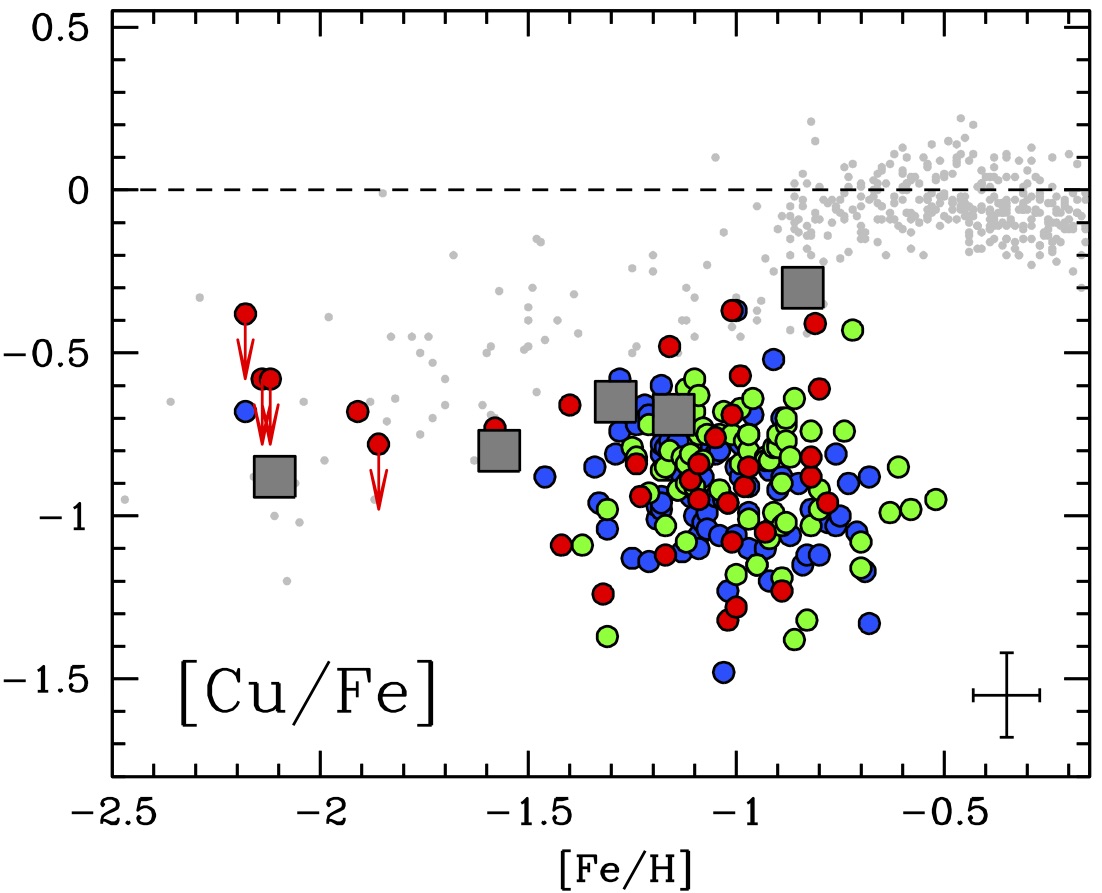}\\
\caption{Behaviour of the iron-elements [Sc/Fe], [V/Fe], [Ni/Fe] and [Cu/Fe] abundance ratios as a function of [Fe/H]. 
Abundances of Galactic field stars are from 
\cite{Reddy2003, Reddy2006, Roederer2014} for all the elements, 
\cite{Gratton2003} for Sc, V and Ni, 
\cite{Fulbright2000} for V and Ni, 
\cite{Adibekyan2012} for Sc and Ni,
\cite{Edvardsson1993, Stephens2002, Bensby2005} for Ni,
\cite{Bihain2004, Yan2015} for Cu.
Same symbols of Fig.~\ref{alfa1}.}
\label{iron1}
\end{figure*}

\begin{figure}[htbp]
\includegraphics[scale=0.225]{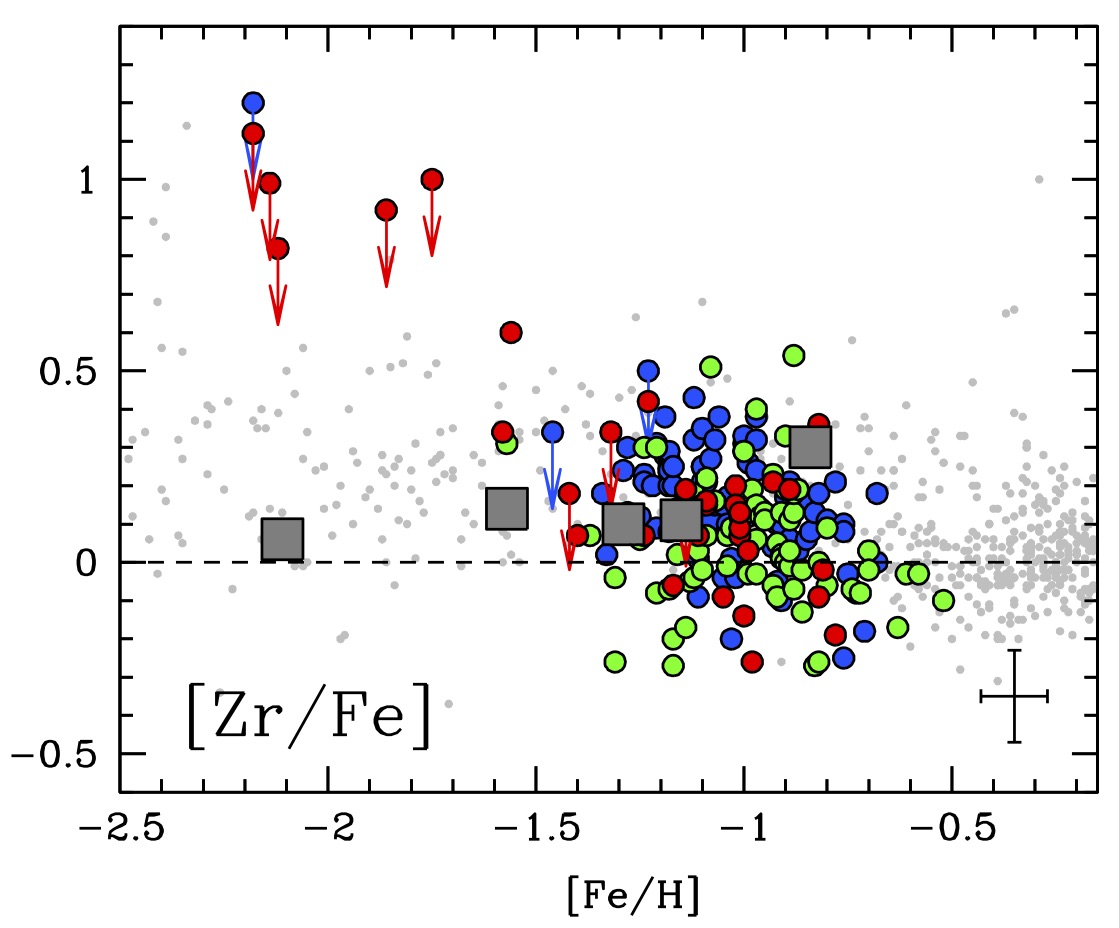}
\includegraphics[scale=0.225]{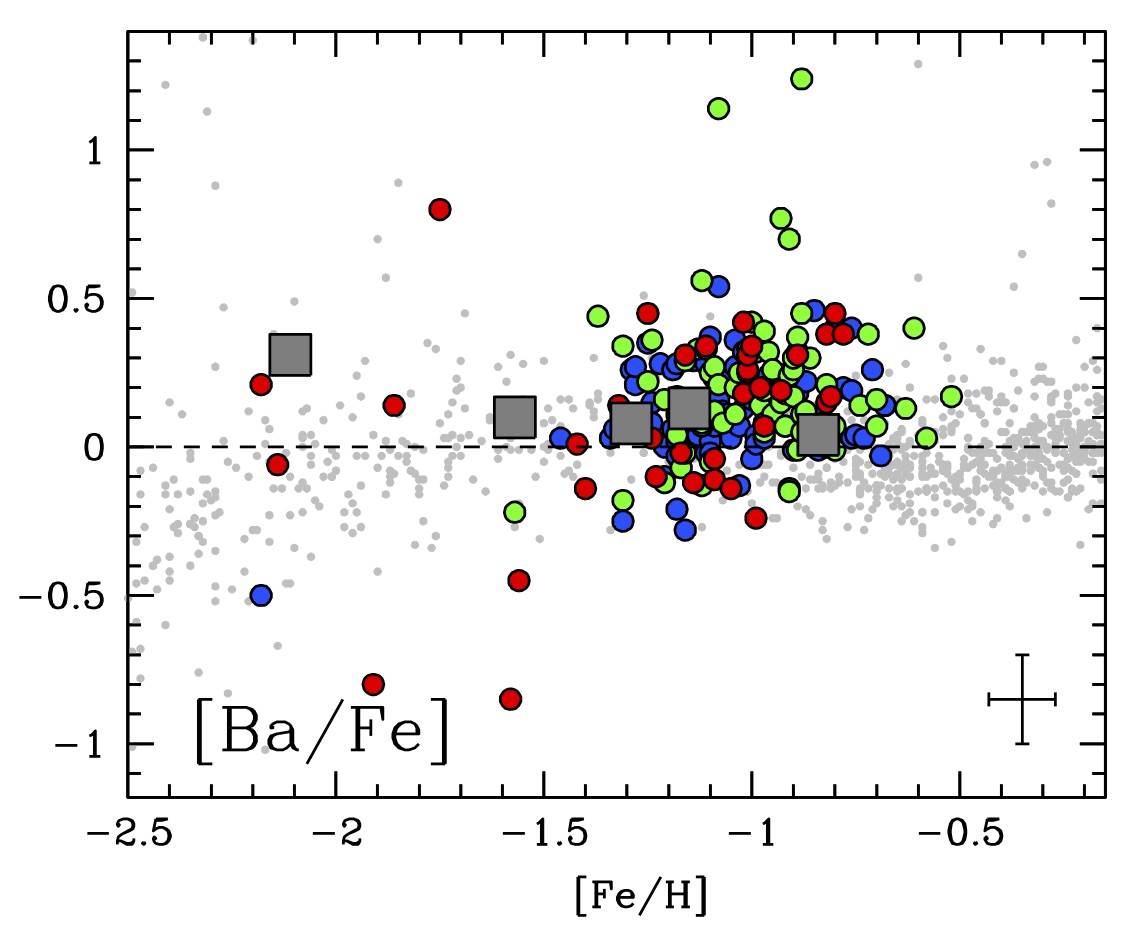}
\includegraphics[scale=0.225]{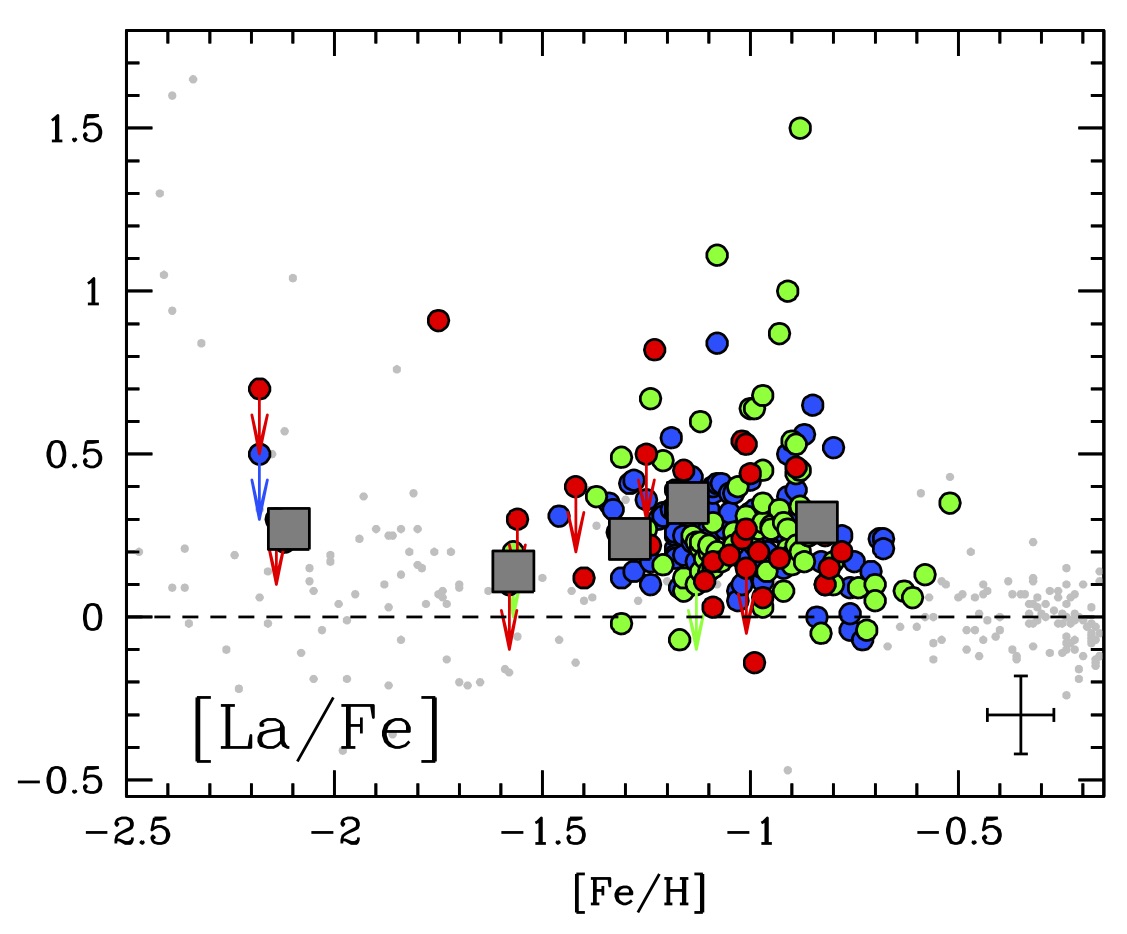}
\caption{Behaviour of the neutron capture-elements 
[Zr/Fe], [Ba/Fe] and  [La/Fe] abundance ratios as a function of [Fe/H]. 
Abundances of Galactic field stars are from 
\cite{mishenina2013, Roederer2014} for all the elements, \cite{Edvardsson1993, Fulbright2000, Reddy2003} for Zr and Ba, 
\cite{Burris2000, battistini16} for Zr and La, \cite{Stephens2002, Barklem2005,  Bensby2005} for Ba.}
\label{slow1}
\end{figure}

Sc and V are produced mainly in massive stars, with a small contribution by SN~Ia only for V \citep{kobayashi20b}.
The SMC stars with [Fe/H]$<$--1.5 dex have Sc and V abundances compatible with those measured in the control sample. 
We note some offsets between these abundance ratios in the control sample and in the literature data, likely 
attributable to different linelists (in terms of log~gf and/or hyperfine structures.
On the other hand, the metal-rich SMC stars have abundances of Sc and V significantly lower than the MW stars (see Fig.~\ref{iron1}). 
For both elements,  we observed a decrease of the abundance ratio with  increasing [Fe/H] because 
of the overwhelming delayed contribution to Fe by SNe~Ia.
This behaviour resembles those observed in metal-rich stars of dwarf galaxies, like LMC and Sagittarius 
\citep[see e.g.][]{sbordone07,minelli21}. We note that also [V/Fe] in metal-rich stars of FLD-419 is systematically 
higher by $\sim$0.15 dex with respect to the stars of FLD-339. A comparable shift has been detected also for [V/Fe] 
in the LMC disc and bar stars \citep{vds13}.

Ni is largely produced by SN~Ia, with production also by CC-SNe, similar to the production of Fe.
The SMC stars have [Ni/Fe] values compatible with those measured in the GCs of the control sample until [Fe/H]$\sim$--1.0 dex, 
while for higher metallicities this abundance ratio slightly decreases, reaching values around [Ni/Fe]$\sim$--0.2 dex (see Fig.~\ref{iron1}). 
A similar behaviour in the SMC stars has been observed by \citet{hasselquist21}.
This mild trend resembles that observed for [Ni/Fe] in the LMC and in Sagittarius at higher [Fe/H] \citep{minelli21}. 
The decrease of [Ni/Fe] at higher metallicities is not observed in MW stars, where [Ni/Fe] remains constant. 
In this respect, \citet{kobayashi20a} suggested a lower 
contribution by sub-Chandrasekhar mass SN~Ia to reproduce the [Ni/Fe] measured 
in dwarf spheroidal galaxies. \\
Cu is produced mainly in massive stars through the weak s-process \citep{romano07}, with a small contribution 
by AGB stars \citep{travaglio04} and a negligible contribution by SN~Ia \citep{iwamoto99,romano07}.
The [Cu/Fe] abundance ratio in the SMC stars exhibits a large star-to-star dispersion and it is difficult to establish 
its real trend. However, it is clear that the most metal-rich SMC stars have [Cu/Fe] lower than 
that measured in MW stars, indicating again a lower contribution to the chemical enrichment 
by massive stars. Values of [Cu/Fe] lower than those measured 
in MW stars have been observed also in the LMC \citep{vds13}, Sagittarius \citep{sbordone07} and Omega Centauri \citep{cunha02}.


\subsection{Neutron capture elements}

Elements heavier than the iron-peak group are produced through neutron capture processes on seed nuclei, 
followed by $\beta$ decays \citep{burbidge57}. The neutron capture elements measured here (namely Zr, Ba and La) 
are produced mainly by the slow process occurring in low-mass (1-3 M$_{\odot}$) AGB stars and in a minor amount 
in more massive stars \citep{busso99,cristallo15}. At low metallicities these elements are produced also through 
rapid processes \citep{truran81}, 
occurring in rare and energetic events like neutron star mergers or collapsars. 
In this spectroscopic dataset, there are no transition 
of pure r-process elements (i.e. Eu) and we cannot discuss the 
relative contribution of these two production channels. 
However, \citet{reggiani21} analysed 4 metal-poor SMC giant stars finding
[Eu/Fe] values higher than those of the MW stars, supporting a strong contribution at these metallicities by r-process.

The SMC stars show [Zr/Fe] and [La/Fe] abundance ratios similar, within the star-to-star scatter, to those observed 
in MW stars, and slightly higher [Ba/Fe]. Generally, these results suggest that the enrichment by AGB stars 
in the SMC has been comparable to that in the MW.
[Ba/Fe] in the SMC stars is enhanced ($\sim$+0.3/+0.4 dex) and higher 
than the values measured in the MW stars. 
[Ba/Fe] displays a large scatter at all the metallicities and not explainable in light of the typical uncertainties 
in the abundance ratios ($\sim$0.15 dex). At [Fe/H]$\sim$--1.0 dex, the SMC stars have values of [Ba/Fe] higher 
than those observed in the MW stars, suggesting a galaxy-wide initial mass function (IMF) biased in favour of the 
low-mass stars in the SMC.
A similar behaviour is observed for [La/Fe], while [Zr/Fe] presents a trend in agreement with that observed for the MW. 
Similar to what we observe for Na and V, also for [Zr/Fe] we found a shift ($\sim$0.2 dex) between FLD-339 and FLD-419 that 
resemble those observed for the same ratio between LMC disc and bar stars \citep{vds13}.
The high values of [Ba/Fa] and [La/Fe], together with the large star-to-star scatter, suggest that the production of s-process elements 
has been very efficient in the SMC, while the large star-to-star scatter could arise from enrichment from AGB stars of different 
metallicities, being the yields of AGB stars for these elements extremely metallicity-dependent.

Finally, we identified a few stars with high [Ba/Fe] and [La/Fe] values ($>$0.5-0.7 dex, reaching also +1.3 dex). 
A similar enhancement of s-process elements could be due to mass transfer processes from an AGB companion star in a binary system.

\section{Conclusions}

The analysis of optical spectra of 206 SMC RGB stars located in three different positions of the parent galaxy 
has allowed us to highlight some finer details of the complex and still poorly known nature of this galaxy. 
The main results are summarised as follows:
\begin{itemize}
\item The RV and [Fe/H] distributions of the three fields are different with each other. 
The fields FLD-339 and FLD-419, despite the same distance from the 
SMC centre, have [Fe/H] distributions peaked at different values, separated by 0.2 dex. 
These two populations could be connected to different bursts of SF 
occurring in the recent life of the SMC \citep{massana22} or the result of a different chemical 
enrichment path in these regions (despite their similar projected distance from the SMC centre).
\item The fraction of metal-poor ([Fe/H]$<$--1.5 dex) stars increases outward, 
being $\sim$1\% in the two internal fields and $\sim$20\% in FLD-121. This run
likely reflects an age gradient in the SMC, with the 
internal regions dominated by intermediate-age, metal-rich stars 
and the outskirts by the old, metal-poor spheroid \citep[see e.g.][]{rubele18}.
\item The RV-[Fe/H] distribution of the observed fields seems to suggest
the possible existence of chemically/kinematic distinct substructures. 
In particular, 
we potentially identified two groups of stars, one around 
[Fe/H]$\sim$--1.1 dex and RV$\sim$+154 \kms\ and the other around 
[Fe/H]$\sim$--0.9 dex and RV$\sim$+172 \kms\ . More data are needed to confirm the statistical 
significance of these chemo-kinematical substructures.
\item The SMC displays, especially for the dominant, metal-rich component, 
distinct abundance patterns with respect to the MW stars. 
In particular, those elements mainly produced by massive stars (Na, $\alpha$, Sc, V and Cu) 
have abundance ratios lower than those measured in the MW stars. 
This suggests that the gas from which these stars formed has
been poorly enriched by the most massive stars. 
This can be explained in light of the low SF rate expected for a galaxy as small as the SMC, leading to a
lower contribution by massive stars to the overall chemical enrichment of the galaxy \citep{jer18,yan20}. 
This is confirmed also by the most metal-poor stars of the sample that exhibit 
[O/Fe] and [Mg/Fe] ratios slightly lower than those in MW stars of similar [Fe/H]. 
\item The [s/Fe] abundance ratios are enriched with respect to the MW stars, 
with a large star-to-star scatter, suggesting that these elements are produced by AGB stars of different masses and 
metallicities. Also, the enhancement of the [s/Fe] abundance ratios in the SMC seems to suggest a galaxy-wide IMF
biased in favour of the low-mass stars in the SMC.
\item 
The possibility that the IMF is not universal, but varies with the environment is the subject of lively debate 
\citep{bastian10,hopkins18,smith20}. 
Theoretically, if stars form in clusters according to IMFs that depend on the metallicity and density 
of the parent gaseous clumps, it is possible to calculate the integrated galaxy-wide IMF that in turn depends 
on the metallicity and star formation rate of the host galaxy 
\citep{jer18,yan20}. 
Moreover, the abundance ratios of chemical elements produced in stars with initial masses falling in narrow and well-detached 
ranges can be used as powerful, indirect probes of the shape of the galaxy-wide IMF \citep[e.g.][]{romano17}. 

Observationally, the possibility that the Sagittarius dwarf spheroidal galaxy had a stronger contribution from AGB stars 
to its chemical enrichment than the MW and the LMC is discussed in \citet{hasselquist21}. 
Similarly, \citet{hallakoun21}, resting on Gaia DR2 data, point to a bottom-heavy IMF for the Gaia-Enceladus progenitor. 
Finally, \citet{muc21} claim that the LMC GC NGC~2005 must have formed in an accreted system that experienced an extremely 
low star formation rate and, 
hence, an extremely low number of hypernova explosions, 
in order to explain the peculiarly low Zn abundance of the cluster. 

On the other hand, \citet{hill19} do not find any clear cut evidence in favour of a non-standard IMF in the Sculptor 
dwarf spheroidal galaxy.  In a forthcoming paper, we will quantitatively deal with the issue of IMF variations in the SMC 
by computing chemical evolution models specifically tailored to this galaxy (Romano et al., in preparation).

\item The three fields exhibit similar chemical patterns for all the elements but  
Na, V, Zr and Ti showing subtle differences among the fields. 
Differences in the same abundance ratios have been observed 
also in the LMC between bar and disk stars \citep{vds13}. 
These differences confirm that the chemical enrichment history in the SMC has been 
not uniform but depends on the position within the galaxy.
\end{itemize}

These promising results enforces the need to study the properties of the SMC stars locally rather than globally, 
with an effort to enlarge the samples of high-resolution spectra 
located in different regions of the galaxy.
In this respect, the advent of the multi-object spectrographs 
like MOONS at the Very Large Telescope \citep{cirasuolo} 
and 4MOST at the VISTA Telescope \citep{dejong19}
will allow us a significant improvement in the investigation of possible chemically-distinct sub-structures 
in the Magellanic Clouds \citep{gonzalez}.

\begin{acknowledgements}
We thanks the referee, Mathieu Van der Swaelmen, for the 
useful comments and suggestions.
This research is funded  by the project "Light-on-Dark" , granted by the Italian MIUR
through contract PRIN-2017K7REXT.
C. Lardo acknowledges funding from Ministero dell'Università e della Ricerca through the Programme {\em Rita Levi Montalcini} (grant PGR18YRML1).
This work has made use of data from the European Space Agency (ESA) mission {\it Gaia} (\url{https://www.cosmos.esa.int/gaia}), 
processed by the {\it Gaia} Data Processing and Analysis Consortium (DPAC,\url{https://www.cosmos.esa.int/web/gaia/dpac/consortium}). 
Funding for the DPAC has been provided by national institutions, in particular the institutions participating in the {\it Gaia} Multilateral Agreement.

\end{acknowledgements}

{}

\end{document}